# A Framework for Understanding the Patterns of Student Difficulties in Quantum Mechanics

Emily Marshman and Chandralekha Singh

*Department of Physics and Astronomy, University of Pittsburgh, Pittsburgh, PA, 15260, USA*

**Abstract**: Compared with introductory physics, relatively little is known about the development of expertise in advanced physics courses, especially in the case of quantum mechanics. Here, we describe a framework for understanding the patterns of student reasoning difficulties and how students develop expertise in quantum mechanics. The framework posits that the challenges many students face in developing expertise in quantum mechanics are analogous to the challenges introductory students face in developing expertise in introductory classical mechanics. This framework incorporates both the effects of diversity in upper-level students' prior preparation, goals, and motivation in general (i.e., the facts that even in upper-level courses, students may be inadequately prepared, have unclear goals, and insufficient motivation to excel) as well as the "paradigm shift" from classical mechanics to quantum mechanics. The framework is based on empirical investigations demonstrating that the patterns of reasoning, problem-solving, and self-monitoring difficulties in quantum mechanics bear a striking resemblance to those found in introductory classical mechanics. Examples from research in quantum mechanics and introductory classical mechanics are discussed to illustrate how the patterns of difficulties are analogous as students learn to unpack the respective principles and grasp the formalism in each knowledge domain during the development of expertise. Embracing such a framework and contemplating the parallels between the difficulties in these two knowledge domains can enable researchers to leverage the extensive literature for introductory physics education research to guide the design of teaching and learning tools for helping students develop expertise in quantum mechanics.

## I. INTRODUCTION

A solid grasp of the fundamental principles of quantum physics is essential for many scientists and engineers. However, quantum physics is a technically difficult and abstract subject. The subject matter makes instruction quite challenging, and even capable students constantly struggle to develop expertise and master basic concepts.

In order to help students develop expertise in quantum mechanics, one must first ask how experts compare to novices in terms of their knowledge structure and their problem-solving, reasoning, and metacognitive skills. According to Sternberg [1], some of the characteristics of an expert in any field include: 1) having a large and well organized knowledge structure about the domain; 2) spending more time in determining how to represent problems than searching for a problem strategy (i.e., more time spent analyzing the problem before implementing the solution); 3) working forward from the given information in the problem and implementing strategies to find the unknowns; 4) developing representations of problems based on deep, structural similarities between problems; 5) efficient problem-solving; when under time constraints, experts solve problems more quickly than novices, and 6) accurately predicting the difficulty in solving a problem. Additionally, experts are more flexible than novices in their planning and actions [2].

Experts also have more robust metacognitive skills than novices. Metacognitive skills, or self-regulatory skills, refer to a set of activities that can help individuals control their learning [3]. The three main self-regulatory skills are planning, monitoring, and evaluation [4]. Planning involves selecting appropriate strategies to use before beginning a task. Monitoring is the awareness of comprehension and task performance. Evaluation involves appraising the product of the task and re-evaluating conclusions [4]. Self-regulatory skills are crucial for learning in knowledge-rich domains. For example, in physics, students benefit from approaching a problem in a systematic way, such as analyzing the problem (e.g., drawing a diagram, listing knowns/unknowns, and predicting qualitative features of the solution that can be checked later), planning (e.g., selecting pertinent principles/concepts to solve the problem), and evaluating (e.g., checking that the preceding steps are valid and that the answer makes sense) [5]. When experts repeatedly practice problems in their domain of expertise, problem-solving and self-regulatory skills may even become automatic and subconscious [3]. Therefore, unless experts are given a new, "novel" problem, they may go through the problem-solving process automatically without making a conscious effort to plan, monitor, or evaluate their work [5, 6].

How can a student become an expert in physics, whether at an introductory or advanced level? There is a vast amount of research literature focusing on student reasoning difficulties in introductory courses, how students in introductory courses differ from physics experts in their problem-solving and self-regulatory skills, and the strategies that may help students become better problem solvers and independent learners (e.g., see Refs. [7-9]). Relatively few investigations have focused on the nature of expertise of advanced physics students and strategies that can be used in upper-level courses to help them build a robust knowledge structure and develop their problem-solving, reasoning, and metacognitive skills [10-17].

Investigations on the nature of expertise development in upper-level courses can benefit from having a framework, even if rudimentary, on which to develop research studies and interpret results. The framework can be refined further as more empirical evidence becomes available. Here, we describe a framework for understanding patterns of student reasoning difficulties and how students develop expertise in quantum mechanics. The framework proposes that the challenges many students face in developing expertise in upper-level quantum mechanics are analogous to the challenges students face in developing expertise in introductory classical mechanics. These analogous patterns of difficulties are often associated with the diversity in the goals, motivation, and prior preparation of upper-level students (i.e., the facts that even in an upper-level physics course, students may be inadequately prepared, have unclear goals, and may not have sufficient motivation to excel) as well as the "paradigm shift" from classical mechanics to quantum mechanics. The framework is based on research demonstrating that the patterns of difficulties in the context of quantum mechanics bear a striking resemblance to those found in introductory classical mechanics.

Why is it useful to have a framework for understanding the patterns of student difficulties and how students develop expertise in quantum mechanics? One common assumption of many physics instructors is that a majority of upper-level physics students are like them, having developed significantly better problem-solving, reasoning, and metacognitive skills than students in introductory physics. Instructors may also presume that, even without guidance and scaffolding support, upper-level students will automatically focus on building a robust knowledge structure of physics. In particular, many instructors assume that most upper-level physics students have developed good learning strategies, are eager and "primed" to learn in all their courses, and are unlikely to struggle in the same manner as students in introductory courses. However, research suggests that there is a large diversity in the preparation of students even in upper-level courses, both in terms of students' content knowledge and their problem-solving and self-regulatory skills [18, 19]. If an instructor of an upper-level course targets instruction at a certain level, many underprepared students will struggle to learn. Furthermore, students have various motivations and goals for enrolling in a course and what they want to get out of a course. Many students will not necessarily be able to learn if the level of instruction is too advanced based on their current knowledge state. This problem is likely exacerbated in a traditionally taught course which does not accommodate the inadequate prior preparation of students and mainly involves lectures which are targeted assuming a certain level of expertise. Moreover, classical mechanics and quantum mechanics are two significantly different paradigms. Therefore, learning quantum mechanics can be challenging even for students who have developed a good knowledge structure of classical mechanics [20, 21]. Adopting such a framework and contemplating the analogous patterns of student difficulties in quantum mechanics and introductory classical mechanics can aid researchers to utilize the extensive literature about introductory physics education in the design of teaching and learning tools for helping students develop expertise in quantum mechanics.

In the following sections, we first give an overview of the framework. We discuss why the diversity in the student population (which implies that there are students with inadequate preparation, unclear goals, and insufficient motivation for excelling in the course) and the novel nature of the quantum paradigm make learning challenging in ways that are analogous to the challenges introductory students face in developing expertise in classical mechanics. Then, we describe how introductory physics students and upper-level students may face similar patterns of difficulty as they learn to unpack the respective principles and grasp the formalism in each knowledge domain during the development of expertise. Furthermore, we discuss empirical research data that provide evidence to support the framework and use concrete examples to illustrate how the patterns of student reasoning, problem-solving, and self-monitoring difficulties are analogous in these two sub-domains of physics. We also discuss how students' inadequate preparation, unclear goals, and insufficient motivation along with the paradigm shift can result in, e.g., a lack of a robust knowledge structure and effective problem-solving skills, transfer difficulties, lack of self-regulation, cognitive overload, and unproductive epistemologies. The concluding section focuses on the implications of this framework for quantum mechanics instruction and research-based instructional design. We discuss how the analogous patterns of difficulties in the two sub-domains of physics can inspire suitable adaptation of research-based strategies. In particular, research-based strategies for helping students develop expertise in introductory mechanics may also be effective in helping upper-level students learn quantum mechanics better.

## II. OVERVIEW OF THE FRAMEWORK

The framework for understanding patterns of difficulties and the development of expertise in quantum mechanics posits that the challenges many students face in upper-level quantum mechanics are analogous to the challenges introductory students face in classical mechanics. Figure 1 summarizes how the increased diversity in the student population, which implies that students who enroll in a course do not necessarily have adequate prior preparation, clear goals, and sufficient motivation to excel, combined with the "paradigm shift" can result in analogous patterns of learning difficulties in introductory mechanics and quantum mechanics. These factors can lead to difficulties in building a robust knowledge structure, developing effective problem-solving, reasoning, and metacognitive skills, transferring knowledge from one context to another, managing cognitive load, and developing productive epistemological views in each of these sub-domains of physics.

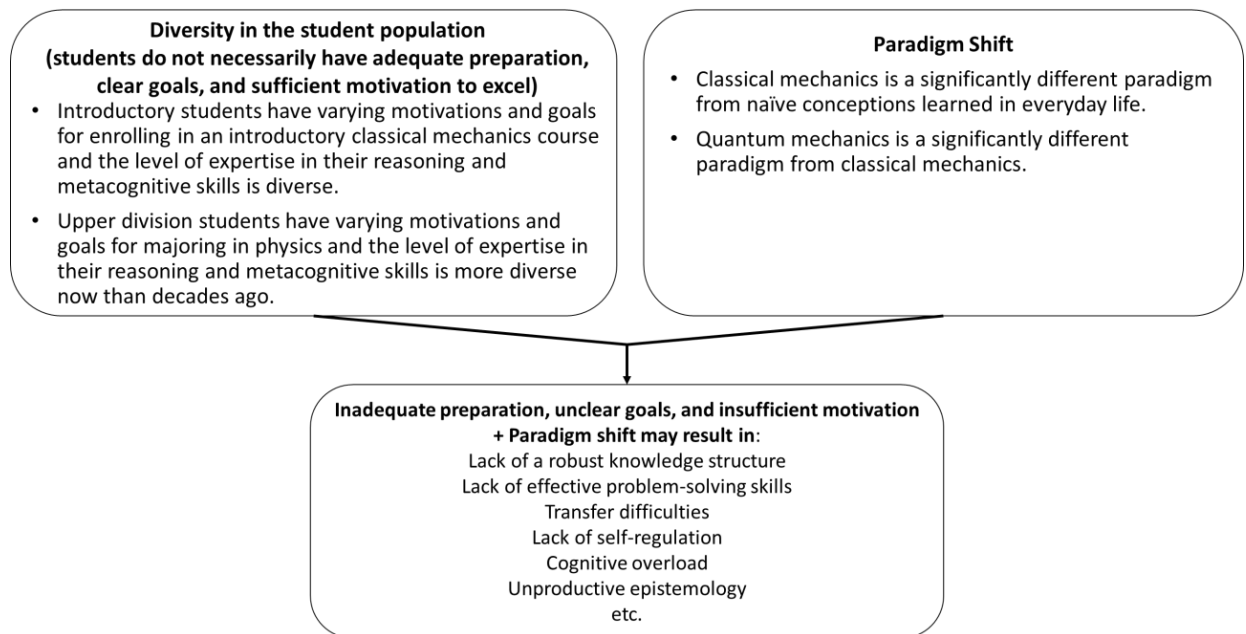

*Figure 1. Framework for understanding why patterns of difficulties in quantum mechanics are analogous to those in introductory classical mechanics.*

### A. Diversity in students' prior preparation, goals, and motivation

Introductory physics is highly abstract and requires logical problem-solving, formal reasoning, and mathematical skills [22]. Many students have not mastered these types of skills by the time they enroll in an introductory college physics course and face difficulties in developing expertise. McDermott points out that the introductory student cannot be thought of as a "younger version" of the instructor [26]. She says that traditional introductory physics courses worked well for instructors, as they do for typically only 1 out of every 30 students in the class [23]. She emphasizes that "a large number of introductory students are inadequately prepared for the level of instruction. Unfortunately, a disproportionate percentage of minority students falls into this category" ([23], p. 302). Halloun and Hestenes developed a composite index, called the competence index, which is determined by students' prior preparation in physics and mathematics (as determined by the performance on diagnostic tests in physics and mathematics administered at the beginning of the course) and showed that the competence index has a significant correlation with students' performance at the end of the course [34]. Based upon the competence index, they state that "with probabilities greater than 0.60 in the large student population we have studied, high competence students were likely to receive an A or B course grade, average competence students were likely to receive a C grade, and low competence students were likely to receive a D or E grade" in traditionally taught algebra or calculus-based introductory physics courses ([34], p. 1047).

In fact, between the years 2003-2009, approximately 42% of beginning postsecondary students took remedial coursework in mathematics [24]. In 2000, approximately 20% of freshmen intending to major in science and engineering reported needing remedial work in mathematics and approximately 10% of them reported needing remedial work in science [24]. These percentages increase for women and minorities—of all freshmen science and engineering majors, approximately 26% of women and 40% of minorities reported needing remedial work in math in 1995 [25]. Students also have various goals and motivations for taking a physics course. They may enroll in a physics course because it is required for their major or they may have an intrinsic interest in the subject. Majors for students taking introductory physics include, for example, computer and information science, biology, medicine, mathematics, chemistry, and engineering. Students in these majors are likely to have diverse goals which can affect their motivation to develop a coherent knowledge structure of introductory physics.

Similar to the diversity of introductory students which makes teaching and learning challenging as McDermott, Halloun and Hestenes point out [23, 26, 34], there is also considerable diversity in upper-level students' preparation, motivation, and goals. Prior investigations have shown that there is a large diversity in both the content knowledge and in the problem-solving, reasoning, and self-regulatory skills of upper-level physics students in quantum mechanics [18, 19]. The goals and motivations for majoring in physics and the preparation of students in upper-level physics courses have gradually become more diverse [27]. A variety of statistics available from the American Institute of Physics (AIP) on undergraduate and graduate education point to the diverse goals and motivations of students enrolling in physics courses [27]. According to AIP data, the percentage of physics Ph.D. students pursuing an academic career (including all types of post-secondary institutions) has steadily decreased over time to approximately 20% currently [27]. AIP data also show that upper-level students' career plans have become more diverse in recent decades, which can impact their motivation to engage deeply with the material [27].

On the other hand, instructors of upper-level physics courses often assume that a majority of their students have already developed robust problem-solving, formal reasoning, mathematical, and self-regulatory skills. They may also believe that upper-level students will automatically make an effort to build a robust knowledge structure, engage in sense-making, and learn from their mistakes without guidance and scaffolding support. Instructors may also assume that all students have goals similar to their own when they were students and are intrinsically motivated to learn. Thus, instructors may teach the way they were taught, i.e., using the traditional approach, assuming that all students are "primed" to learn. Most do not take into account the diversity in students' prior preparation, goals, and motivation. This traditional approach is in contrast to the central tenets of PER-based instructional approaches, which focus on in-class and out-of-class activities and self-paced tools to build on the prior knowledge of a diverse group of students to help them develop expertise.

There is no doubt that some upper-level students are well prepared, have clear goals, and are sufficiently motivated to excel. While the percentage of such students in an upper-level course may be more than 1 out of 30 students as in an introductory physics course [23], a significant portion of students even in an upper-level course are neither intrinsically motivated to learn physics like their instructors nor are prepared or "primed" to learn from a traditional "lecture only" approach [18, 19]. This situation is similar to students in introductory physics courses failing to learn from traditional lectures alone [21, 28, 29]. Figure 2 summarizes the connections between the diversity in students' backgrounds and the final state of expertise in a traditional course (either introductory classical mechanics or quantum mechanics). For an individual student at the beginning of instruction, his/her prior preparation, goals, and motivation (PGM) can be thought of as components which can be weighted appropriately for an individual student to yield a composite PGM score (or "PGM," in short) to determine where he/she initially falls on the PGM spectrum (see Figure 2). Similar to each student's composite PGM in introductory physics, each student's PGM in an upper-level course can affect the extent and manner in which the student engages while learning quantum mechanics. In both introductory and upper-level courses, there is a distribution of individual student's PGM scores. Some students are highly prepared, motivated, and have clear goals while others are underprepared, have unclear goals, and lack sufficient motivation to excel. Traditional instruction may only benefit students above a threshold PGM at which instruction is targeted. In fact, highly prepared and motivated upper-level students may become experts in quantum mechanics regardless of the type of instruction. Some students who are not adequately prepared but are motivated to learn and have clear goals may also manage to develop expertise even in a traditionally taught course. However, underprepared students lacking clear goals and motivation who fall below the threshold PGM score (based upon the level at which the instruction is targeted in a traditional course) will struggle to develop expertise in a traditionally taught physics course that does not take into account individual student's prior knowledge and builds on it. These types of students may display learning difficulties which are analogous to the difficulties displayed by students learning introductory classical mechanics.

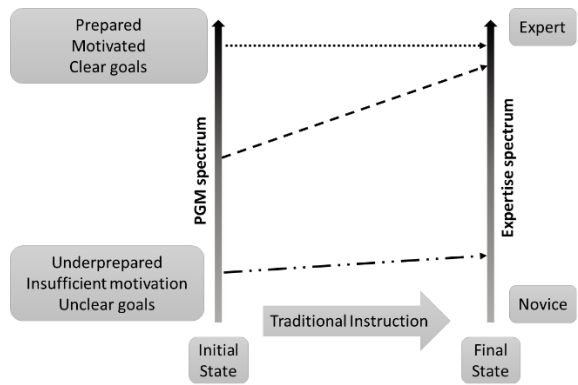

*Figure 2. A student's initial preparation, goals, and motivation, when weighted appropriately, can yield a composite PGM score (or PGM). If a student's PGM is below a certain threshold, it can result in learning difficulties and impact the student's path toward expertise, especially in a traditionally taught course (either introductory classical mechanics or quantum mechanics). The schematic diagram shows the effect of three students' PGM scores on their transition from an initial state to a final state in introductory classical mechanics or quantum mechanics. Dashed lines schematically represent students' paths toward expertise in a traditionally taught course.*

### B. The paradigm shift

While the diversity in students' preparation, goals, and motivations (specifically, the fact that there are students with a PGM score below a certain threshold at which instruction is targeted) may partly account for the difficulties in learning classical mechanics and quantum mechanics, difficulties may be exacerbated by the fact that the paradigms of classical mechanics and quantum mechanics are significantly different than the paradigms that students in the respective courses have previously learned. Therefore, the very nature of a new paradigm causes additional difficulties for students. In his book *The Structure of Scientific Revolutions*, Thomas Kuhn focuses on the concept of a "paradigm shift," i.e., how insurmountable problems lead scientists to question a traditional paradigm's assumptions and a new paradigm emerges [30]. He states that "the reception of a new paradigm often necessitates a redefinition of the corresponding science" ([30], p. 103). He discusses how new paradigms are born from older paradigms and, as such, they often incorporate elements such as vocabulary, concepts, and experiments of the prior paradigm. However, the new paradigm does not utilize these elements in the traditional way which can result in misunderstandings between the two paradigms and lead to learning difficulties. Kuhn discusses the example of the paradigm shift from classical mechanics to quantum mechanics to demonstrate how difficult it is to reconcile the old and new paradigms. He notes that as individuals begin learning a new paradigm, they may continue to apply their knowledge of the older paradigm onto the new paradigm, and this is not surprising. He states that, at least partly, "the source of resistance is the assurance that the older paradigm will ultimately solve all its problems, that nature can be shoved into the box the paradigm provides" ([30], pp. 151-152).

Kuhn's work influenced science education research and inspired the theory of conceptual change [31, 32]. Research suggests that introductory physics students constantly try to make sense of the world around them. The mental models they build of how things work in everyday life are based on naïve reasoning and limited expertise and are often inconsistent with the laws of physics [33]. Moreover, everyday terms such as velocity, acceleration, momentum, energy, work, etc. do not have the same precise meaning as in physics and students must learn to differentiate between how those terms are used in physics vs. how they are used in everyday life. Students in introductory physics must shift from their adherence to their naïve mental models to the models consistent with the new paradigm of classical mechanics. Clement notes that students' resistance in shifting from their naïve mental models to the classical mechanical model is not surprising, since "pre-Newtonian concepts of mechanics had a strong appeal, and scientists were at least as resistant to change as students are" ([22], p. 70). Similarly, McDermott emphasizes that "the student mind is not a blank slate on which new information can be written without regard to what is already there. If the instructor does not make a conscious effort to guide the student into making the modifications needed to incorporate new information correctly, the student may do the rearranging. In that case, the message inscribed on the slate may not be the one the instructor intended to deliver" ([23], p. 305). Halloun and Hestenes also note that each student possesses beliefs and "common sense" intuitions about physical phenomena that are derived from their personal experience and they use these "common sense theories" to interpret what is taught in a physics course [34]. In fact, Clement emphasizes that a student possessing robust mathematical skills can "mask his or her misunderstanding of underlying qualitative concepts" ([22], p. 66). Having robust problem-solving skills does not guarantee success in developing a conceptual understanding of introductory physics—the students must revise their own "common sense theories" and build a coherent knowledge structure [22, 23, 34].

Similarly, students learning quantum mechanics must shift their adherence from the concepts and principles learned in classical mechanics to the new quantum paradigm in order to predict and explain quantum phenomena. Because the quantum mechanics paradigm is radically different from the classical paradigm, students must build a knowledge structure for quantum mechanics essentially from scratch, even if they have built a robust knowledge structure of classical mechanics. It is true that students are unlikely to have unproductive mental models about quantum mechanics concepts before formal instruction in quantum mechanics because one does not routinely encounter situations that require reasoning about quantum processes in everyday life. Therefore, one might assume that learning quantum mechanics may be easier than classical mechanics in this regard. However, as Kuhn suggests, the physics content knowledge that students learned in earlier courses, including classical mechanics, can interfere with building a robust knowledge structure of quantum mechanics. Similar to the possibility of naïve notions about velocity, momentum, or work from everyday experience interfering with learning classical mechanics, concepts of position, momentum, angular momentum, etc. are embedded so differently in the classical mechanics and quantum mechanics formalisms that intuition about these concepts developed in classical mechanics can actually interfere with learning quantum mechanics. For example, in quantum mechanics, the connection between quantum formalism and phenomena is made through measurement and inferences about physical observables, e.g., position, momentum, energy, and angular momentum. But unlike classical mechanics, a particle does not, in general, have a definite position, momentum, or energy in quantum mechanics. In quantum mechanics, all information about a system is contained in the state vector or wave function which lies in an abstract vector space. The measurement of an observable collapses the wave function to an eigenstate of the operator corresponding to the observable measured, and the probability of measuring a particular value can be calculated from the knowledge of the wave function. These novel quantum concepts have no analogs in classical mechanics even though position, momentum, energy, angular momentum, etc. are common terms in both paradigms. Similar to introductory students' difficulties, upper-level students' difficulties with quantum concepts can be masked if they have developed robust mathematical skills. Indeed, the gap between conceptual and quantitative learning can continue to get wider at the advanced level in the traditional mode of instruction that focuses on "plug and chug" approaches to teaching and assessment. Unless upper-level students construct a coherent knowledge structure of quantum mechanics, difficulties at the conceptual level will persist.

### C. Analogous patterns of difficulty in the development of expertise in classical mechanics and quantum mechanics

As discussed in the previous sections, in both introductory physics and quantum mechanics, the large diversity in students' goals, motivations, and prior preparation (specifically, a sufficient number of students with a PGM score below a threshold at which instruction is targeted) coupled with the paradigm shift can result in learning difficulties as students develop expertise in each of these sub-domains of physics. As introductory and upper-level students start to build a knowledge structure about classical mechanics and quantum mechanics, respectively, their knowledge will initially be in disconnected pieces [33, 35] and their reasoning about their respective domains will only be locally consistent and lack global consistency. In fact, there is nothing unusual about students going through this stage. Those who begin their pursuit of developing expertise in any knowledge-rich domain must go through a phase in which their knowledge is in small, disconnected pieces which are only locally consistent, and this "knowledge in pieces" phase causes reasoning difficulties [33, 35, 36-38]. While students struggle to manage many small, disconnected pieces of knowledge, they can experience cognitive overload [39] and may not have the cognitive capacity to engage in self-regulatory activities. Additionally, students may possess relevant knowledge to solve a problem, but they may not invoke or apply relevant knowledge pieces appropriately in certain contexts. Cognitive overload and failure to invoke or apply relevant knowledge may lead to inconsistent reasoning and difficulties in the transfer of knowledge. Each student must go through the process of gradually building a knowledge structure and pass through the "knowledge in pieces" phase [33] while learning classical mechanics and quantum mechanics separately because the conceptual paradigms are sufficiently different in these sub-domains of physics as discussed (even though the same terminology is used, e.g., momentum, energy, etc.).

Instructors of upper-level courses may inadvertently teach students at a level which is not aligned with many students' prior preparation, goals, and motivation because they often assume that a majority of their students have already developed robust problem-solving, reasoning, and metacognitive skills and that they are intrinsically motivated to learn. However, if instruction in quantum mechanics is not aligned with students' prior preparation, many students will struggle to learn because they may not have developed robust problem-solving, reasoning, and metacognitive skills. Furthermore, students have various goals for enrolling in a quantum mechanics course and many of them are not necessarily intrinsically motivated to learn [19]. Even if students are prepared, have clear goals, and are intrinsically motivated to learn, they may bring to bear prior classical conceptions within the new paradigm of quantum mechanics. Thus, students' mastery of classical mechanics does not imply that they will be able to master quantum mechanics without a conscious effort on the part of the students to build a knowledge structure of quantum mechanics and assimilate and accommodate new ideas (and make lateral connections between the classical mechanics and quantum mechanics schema to understand the differences between these formalisms explicitly and when and how they come together, e.g., by taking the classical limit). Therefore, students learning classical mechanics and quantum mechanics are likely to show similar patterns of reasoning difficulties as they move up along the expertise spectrum in each of these sub-domains of physics. In each case, if students continue their efforts to repair, reorganize, and extend their knowledge structure [36-38] they will reach a point where their knowledge structure becomes robust enough that they become a nominal expert. Then, they will be able to make predictions and inferences which are globally consistent within the respective formalisms and their reasoning difficulties will be significantly reduced. Even after becoming a nominal expert, a student's expertise in the respective sub-domain of physics can keep evolving. In the knowledge schema of classical mechanics or quantum mechanics, the strengthening of nodes and building of additional links between nodes (even if there are redundancies in the links) can make students transition from nominal to adaptive experts who can solve more complex problems [36-38, 40].

### III. EXAMPLES OF ANALOGOUS PATTERNS OF STUDENT DIFFICULTIES IN QUANTUM MECHANICS AND INTRODUCTORY CLASSICAL MECHANICS

Upper-level physics students typically display expert-like behavior when solving problems in introductory classical mechanics because they possess a large amount of compiled knowledge about introductory physics due to repetition of the basic content in various courses. They may not need to do much self-regulation while solving introductory problems [36, 41]. However, they may fail to use these skills when solving problems in the domain of quantum mechanics in which they are not experts and may display patterns of difficulties analogous to those of students learning introductory classical mechanics.

Below, we discuss empirical evidence for the framework based upon research on student difficulties in quantum mechanics and introductory classical mechanics. In particular, we discuss concrete examples of difficulties involving: 1) categorization of problems based upon how they are solved; 2) not using problem solving as an opportunity for learning; 3) inconsistent and/or context dependent reasoning; 4) inappropriate or negative transfer from one situation to another; 5) lack of transfer of knowledge; 6) "gut-feeling" responses inconsistent with the laws of physics; 7) solving multi-part problems; and 8) epistemological issues. Each of these types of difficulties are symptoms of ineffective problem solving, a lack of self-regulation, an incoherent knowledge structure, cognitive overload, an inability to transfer knowledge appropriately, or unproductive epistemologies. It is impossible to disentangle the contributions of the paradigm shift and the inadequate preparation, unclear goals, and insufficient motivation of students to each example of difficulty discussed below. However, each difficulty is an indication of how these factors can result in impediments to learning. We note that many of the examples of student difficulties in introductory classical mechanics and quantum mechanics discussed below could be placed into multiple categories of difficulties, but we have typically chosen to place them in one of the categories mentioned earlier since they are used to illustrate a particular type of analogous difficulty in introductory mechanics and quantum mechanics. In particular, we place each example in one category which clearly represents the difficulty.

### A. Categorization of physics problems

#### 1. Poor categorization of problems in quantum mechanics

Categorization of problems refers to grouping problems together based upon how one would solve the problems. Lin and Singh [18] performed an investigation in which physics professors and students from two junior/senior level quantum mechanics courses were asked to categorize twenty quantum mechanics problems based upon the similarity of the solutions. Students completed the categorization task after instruction in relevant concepts. Professors' categorizations were overall rated higher than those of students by three faculty members who evaluated all of the categorizations blindly (without the knowledge of whether the categories were created by the professors or students). Many students categorized quantum mechanics problems based on the surface features of the problems, such as "infinite square well problem," "free particle problem," or "Stern-Gerlach problem." The scores obtained by the students on the categorization task were more or less evenly distributed with some students scoring similar to the professors while other students scored extremely low.

#### 2. Poor categorization of problems in introductory physics

Chi et al. used a categorization task to assess introductory physics students' expertise in classical mechanics after instruction in relevant concepts [42]. Unlike physics experts who categorized problems based on the physics principles (e.g., conservation of mechanical energy, conservation of momentum, etc.), introductory students categorized problems based on surface features, such as "inclined plane problems" and "pulley problems" [42].

#### 3. Possible causes for poor categorization in quantum mechanics and introductory classical mechanics

Categorizing problems based upon similarity of solution is often considered a hallmark of expertise [42, 43]. The wide distribution in students' performance on the categorization tasks in introductory classical mechanics and quantum mechanics suggests that students are still developing expertise in the respective sub-domains of physics. Students' prior preparation, goals, and motivation as well as the paradigm shift can affect the extent to which students develop expert-like approaches to problem categorization. If students have not developed expertise in the respective sub-domains of physics, they may focus on the "surface features" rather than "deep features" of the problems with a negative impact on their performance in categorizing problems. Furthermore, students' goals and motivations for enrolling in a course can impact the extent to which they develop a coherent knowledge structure of classical mechanics or quantum mechanics and are able to group together problems with differing surface features but equivalent deep features. Students may also incorrectly categorize problems based upon their "common sense theories" in classical mechanics or prior classical mechanics knowledge in quantum mechanics, depending on the type of problem posed.

### B. Not using problem solving as a learning opportunity

#### 1. Not using problem solving as a learning opportunity in quantum mechanics

One attribute of physics experts is that they learn from their own mistakes while solving problems. Mason and Singh [19] investigated the extent to which upper-level students in quantum mechanics learn from their mistakes. In this investigation, they administered four problems in the same semester twice, both on the midterm and final exams in an upper-level quantum mechanics course. The performance on the final exam shows that while some students performed equally well or improved compared to their performance on the midterm exam on a given question, a comparable number performed poorly both times or regressed (i.e., performed well on the midterm exam but performed poorly on the final exam). The wide distribution of students' performance on problems administered a second time points to the fact that many advanced students may not automatically exploit their mistakes as an opportunity for repairing, extending, and organizing their knowledge structure. Mason and Singh [19] also conducted individual interviews with a subset of students to delve deeper into students' attitudes toward learning and the importance of organizing knowledge. They found that many students focused on selectively studying for the exams and did not necessarily look at the solutions provided by the instructor for the midterm exams to learn, partly because they did not expect those problems to be repeated on the final exam and/or found it painful to confront their mistakes. When students were given grade incentives to fix their mistakes on a midterm exam, they did significantly better on similar final exam problems than students who were not given a grade incentive to fix their mistakes on the midterm exam [99].

### 2. Not using problem solving as a learning opportunity in introductory classical mechanics

Yerushalmi et al. [44] investigated the extent to which diagnosing one's own mistakes in multi-part recitation quiz problems (by rewarding students for completing a self-diagnosis task during a following recitation class) helped students in introductory classical mechanics perform better on similar exam problems given later. Students in the three intervention groups diagnosed their mistakes by either 1) using a detailed solution provided by the teaching assistant (TA); 2) having the TA outline the main features of the solutions; or 3) consulting their own books and notes. The students in an equivalent comparison group were not explicitly asked to diagnose their mistakes. It was found that, compared to the comparison group, the performance on challenging follow-up exam problems was 46% better for students who diagnosed their mistakes by consulting their books and notes compared to those who were provided a detailed solution. The students in this intervention group were generally more engaged and struggled more to diagnose their mistakes than those in the intervention group in which the TA provided a detailed solution. The study suggests that introductory students do not use problem solving as a learning opportunity unless they are given an incentive (e.g., via grades) to diagnose their mistakes and become cognitively engaged with the material.

### 3. Possible causes for not automatically using problem solving as a learning opportunity in quantum mechanics and introductory physics

Students' goals, motivation, and prior preparation may affect whether they use problem solving as a learning opportunity without an explicit reward system. Students may not automatically use problem solving as a learning opportunity because they have not necessarily developed robust self-monitoring skills. Furthermore, many introductory students and upper-level students have not become independent learners and they do not necessarily have intrinsic motivation to learn from their mistakes. Also, as students are developing expertise in a new paradigm, they may be in a "knowledge in pieces" phase [33] and may not necessarily have the cognitive capacity to automatically learn from their mistakes.

## C. Inconsistent and/or context-dependent reasoning
### 1. Inconsistent and/or context-dependent reasoning in quantum mechanics
#### a) Inconsistent reasoning about the time dependence of an expectation value of an observable in the context of Larmor precession

Eighty-nine students from multiple universities were asked the following question about the time-dependence of the expectation value of an electron spin component [20]:

*An electron is in a uniform magnetic field B which is pointing in the z-direction. The Hamiltonian for the spin degree of freedom for this system is given by $\hat{H} = -\gamma B \hat{S}_z$ where $\gamma$ is the gyromagnetic ratio and $\hat{S}_z$ is the z component of the spin angular momentum operator. If the electron is initially in an eigenstate of $\hat{S}_x$, does the expectation value of $\hat{S}_y$ depend on time? Justify your answer.*

Some students correctly stated that the expectation value of $\hat{S}_y$ is zero if the initial state is an eigenstate of $\hat{S}_x$. However, they incorrectly claimed that the time dependence of the expectation value of $\hat{S}_y$ is also zero. For example, one interviewed student argued that the expectation value is zero when the initial state is not an eigenstate of the spin component whose time dependence of expectation value is desired. His argument was that all eigenstates of $\hat{S}_x$ are orthogonal to all eigenstates of $\hat{S}_y$ (which is actually not true although the expectation value of $\hat{S}_y$ is zero for the given initial state which is an eigenstate of $\hat{S}_x$). The interviewer reminded him that the eigenstate of $\hat{S}_x$ is only the initial state and he had to find the time dependence of the expectation value of $\hat{S}_y$. The student immediately responded: "I understand that ... [but] since the expectation value of $\hat{S}_y$ is zero in the initial state ... so is its time dependence." In the context of an electron which is initially in an eigenstate of $\hat{S}_x$ in a uniform magnetic field, the student was unable to differentiate between the expectation value and rate of change of the expectation value. Students often display inconsistent reasoning in different contexts, e.g., students may recognize the difference between the expectation value and the rate of change of the expectation value in a particular context but are unable to recognize the difference in another context.

### b) Inconsistency in identifying a quantum state in position representation

Students in quantum mechanics courses often display inconsistent reasoning in their responses to consecutive questions. For example, a conceptual, multiple-choice survey was administered to 39 upper-level students to determine the extent to which they use appropriate problem-solving, reasoning, and self-regulatory skills while solving quantum mechanics problems [45]. In addition, think-aloud interviews were conducted with 23 students to observe how they reasoned about the quantum mechanics problems. On the multiple-choice survey, three consecutive conceptual questions were posed about the quantum mechanical wave function in position representation with and without Dirac notation. In the first question, 90% of the students correctly noted that the position space wave function is $\Psi(x) = \langle x|\Psi\rangle$. However, the second question asked about a generic quantum mechanical operator $\hat{Q}$ acting on the state $|\Psi\rangle$ in the position representation, i.e., $\langle x|\hat{Q}|\Psi\rangle$. Students were told that $\hat{Q}$ was a diagonal operator in position representation. Two of the answer choices were $Q(x)\langle x|\Psi\rangle$ and $Q(x)\Psi(x)$, which are both correct since $\Psi(x) = \langle x|\Psi\rangle$. A student who is self-monitoring would note that the two statements are the same since $\Psi(x) = \langle x|\Psi\rangle$. However, 41% of the students claimed that only one of the answers ($Q(x)\langle x|\Psi\rangle$ or $Q(x)\Psi(x)$) is correct, but not both. In the third question, 36% of the students claimed that $\langle x|\Psi\rangle = \int x\Psi(x)dx$ is correct. However, it is incorrect because if $\Psi(x) = \langle x|\Psi\rangle$, then $\langle x|\Psi\rangle = \int x\Psi(x)dx = \Psi(x)$ does not make sense. In a fourth consecutive question, 39% of the students claimed that $\langle x|\Psi\rangle = \int \delta(x - x')\Psi(x')dx'$ is incorrect (it is a correct equality because the integral results in $\Psi(x)$). In the interview, a graduate student who noted correctly that $\Psi(x) = \langle x|\Psi\rangle$ but who incorrectly claimed that $\langle x|\Psi\rangle = \int \delta(x - x')\Psi(x')dx'$ is incorrect reasoned as follows "… it just doesn't seem correct, that $\Psi(x)$ should just pop out [of the integral]. It's giving you just a wave function of $x$ and I just don't like that. I think [the inner product] should just give you a number." He correctly reasoned that the inner product is a number, but did not make the connection that $\Psi(x)$ is also a number for any particular value of $x$. He was so focused on his concern that the inner product is a number that he did not notice the inconsistency between the responses to this question and question 1 in which he appeared quite confident that $\Psi(x) = \langle x|\Psi\rangle$. We note that the integrals of the type shown above are simple to solve for an advanced student taking quantum mechanics if the problem is given as a math problem without the quantum mechanics context. However, in the context of quantum mechanics, the integral involving a delta function was enough to make this student (and many others) concerned about whether the physical content of that statement made sense from the point of view of quantum mechanics when the integral was nothing more than $\Psi(x) = \langle x|\Psi\rangle$.

### c) Inconsistency in determining possible wave functions for an infinite square well

Research suggests that many students are inconsistent in their responses to whether a specific wave function is allowed for an infinite square well [46]. For example, the following question was administered to students in quantum mechanics courses at many different universities:

*Which of the following wave functions are allowed for an electron in a one dimensional infinite square well of width a with boundaries at $x = 0$ and $x = a$? In each of the three cases, A is a suitable normalization constant. You must provide clear reasoning for each case.*

I.  $A sin^3(\pi x/a)$
II. $A(\sqrt{2/5}\, sin(\pi x/a) + \sqrt{3/5}\, sin(2\pi x/a))$
III. $Ae^{-((x-a/2)/a)^2}$

The wave function $Ae^{-((x-a/2)/a)^2}$ is not allowed because it does not satisfy the boundary conditions. The first two wave functions are both smooth functions which satisfy the boundary conditions, so each can be written as a linear superposition of the stationary states. Therefore, they are both possible wave functions. However, 45% of the students claimed that $A sin^3(\pi x/a)$ is not an allowed wave function but $A(\sqrt{2/5}\, sin(\pi x/a) + \sqrt{3/5}\, sin(2\pi x/a))$ is an allowed wave function. The most common reason for claiming that $A sin^3(\pi x/a)$ is not allowed was that it does not satisfy the time-independent Schrödinger equation, $\hat{H}\Psi(x) = E\Psi(x)$. Students incorrectly claimed that the time-independent Schrödinger equation was the equation that all allowed wave functions should satisfy. Many of the students asserted that $Asin^3(\pi x/a)$ does not satisfy $\hat{H}\Psi(x) = E\Psi(x)$ but $A(\sqrt{2/5}\sin(\pi x/a) + \sqrt{3/5}\sin(2\pi x/a))$ does, which is incorrect (because neither satisfies the time-independent Schrödinger equation). In their reasoning, many students explicitly wrote the Hamiltonian as $(-\hbar^2/2m)\partial^2/\partial x^2$ and showed that the second derivative of $Asin^3(\pi x/a)$ would not yield the same wave function back multiplied by a constant, which is a correct statement (although it does not imply that $Asin^3(\pi x/a)$ is not a possible wave function, which is what they were trying to prove). On the other hand, the same students did not attempt to explicitly take the second derivative of $A(\sqrt{2/5}\sin(\pi x/a) + \sqrt{3/5}\sin(2\pi x/a))$, which also does not yield the same wave function back multiplied by a constant. For this wave function, a majority claimed that it is a possible wave function because it is a linear superposition of the functions $sin(n\pi x/a)$. Many students used incorrect inconsistent reasoning, claiming that a linear superposition of stationary states would satisfy the time-independent Schrödinger equation $\hat{H}\Psi(x) = E\Psi(x)$ but that $Asin^3(\pi x/a)$ would not satisfy it (even though $Asin^3(\pi x/a)$ can also be written as a linear superposition of two stationary states). Furthermore, in interviews, when students were asked whether or not $sin^3(x)$ can be written as a sum of sine functions, some of them remembered that it can be written as a sum of sine functions (i.e., $sin^3(x) = (3\sin(x) - \sin(3x))/4$). But even with this realization, some students did not change their previous answer that $Asin^3(\pi x/a)$ is not a possible wave function but $A(\sqrt{2/5}\sin(\pi x/a) + \sqrt{3/5}\sin(2\pi x/a))$ is. Thus, while many students explicitly showed that $Asin^3(\pi x/a)$ did not satisfy the time-independent Schrödinger equation and recalled that $Asin^3(\pi x/a)$ can be written as a linear superposition of sine functions, they did not use this knowledge to interpret that linear combinations of stationary states with different energies do not satisfy the time-independent Schrödinger equation (contrary to their claim).

d) Inconsistent reasoning about a possible wave function for a finite square well

Another instance of inconsistent reasoning in quantum mechanics is displayed in the following example. The following question about a particle in a finite square well was administered to 226 students from multiple universities [47]:

*Choose all of the following statements that are correct about the wave function for a particle in a one-dimensional finite square well shown below at time $t = 0$. $\Psi(x,0)$ and $d\Psi(x,0)/dx$ are continuous and single-valued everywhere. The wave function $\Psi(x,0)$ is zero in the regions $x < b_1$ and $x > b_2$. Assume that the area under the $|\Psi(x,0)|^2$ curve is 1.*

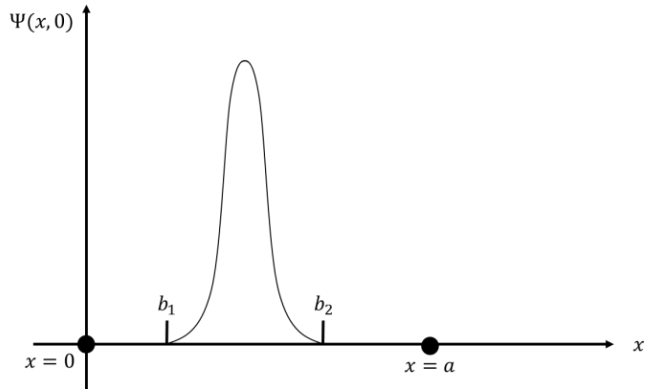

*Figure 3. A possible wave function for a finite square well. Students in an upper-level quantum mechanics course were asked if it is a possible wave function for a finite square well.*

*(1) It is a possible wave function.*
*(2) It is not a possible wave function because it does not satisfy the boundary conditions. Specifically, it goes to zero inside the well.*
*(3) It is not a possible wave function because the probability of finding the particle outside the finite square well is zero but quantum mechanically it must be nonzero.*
*A. 1 only    B. 2 only    C. 3 only    D. 2 and 3 only    E. None of the above*

Forty percent of the students selected the correct response (A). Fifty-five percent of the students incorrectly responded that it is not a possible wave function. Students correctly reasoned that for a finite square well, the particle has a non-zero probability of being in the classically forbidden region in a stationary state. However, they incorrectly overgeneralized this knowledge and claimed that any possible wave function for this system must also have a nonzero probability in the classically forbidden region [47]. In individual interviews, students who answered the above question incorrectly were asked if a highly localized function (approximately a delta function in position) could represent a possible wave function because that is what one obtains after a position measurement. Some students readily responded that a delta function could represent a possible wave function because you can obtain a delta function wave function after a position measurement. However, some of them failed to reason that if a delta function can be a possible wave function for a finite square well, then the wave function in Figure 3 can also be a possible wave function. Students did not note the inconsistency in their statements that a delta function is a possible wave function but the wave function in the question discussed above does not represent a possible wave function for a finite square well. For the wave function shown in Figure 3, students focused on the fact that the stationary state wave functions of a finite square well have non-zero values in the classically forbidden region.

### 2. Inconsistent and/or context-dependent reasoning in introductory classical mechanics

#### a) Inconsistent reasoning about velocity and acceleration

Similar to upper-level students in quantum mechanics claiming that if the expectation value of an operator is zero in an initial state, then the time dependence of the expectation value of that operator is also zero, introductory students often claim that if the velocity of a particle is zero, the particle must have zero acceleration (rate of change of velocity is zero). This type of inconsistent reasoning about a physical quantity and rate of change of that physical quantity has been found in other introductory physics contexts as well.

#### b) Inconsistent reasoning about Newton's Second Law

A lack of consistency in student responses is well-documented in introductory physics. In introductory mechanics, a student may correctly reason in a simple context that a larger net force on an object would imply a larger acceleration (as opposed to a larger constant velocity), but incorrectly claim that the net force is larger on an object moving at a constant velocity of $2\vec{v}_o$ compared to one that is moving at $\vec{v}_o$ [48].

#### c) Inconsistent reasoning in applying Newton's Third Law

On the Force Concept Inventory [48], there are many questions involving Newton's Third Law. Typically, the percentage of students at a typical state university who answer these questions correctly is quite varied, with approximately 80% of the students providing correct responses in some contexts while only approximately 20% provide correct responses for other questions [48].

### 3. Possible causes for inconsistent reasoning and/or context-dependent reasoning in quantum mechanics and introductory classical mechanics

The above examples indicate that students in both introductory classical mechanics and quantum mechanics may fail to use appropriate problem-solving and self-regulatory skills while solving problems in a domain in which they are still developing expertise. This may be due, in part, to students' inadequate prior preparation. Students who have not developed robust problem-solving, reasoning, and metacognitive skills will display inconsistent reasoning while solving problems. Furthermore, the examples illustrate that students learning a new paradigm may discern the applicability of appropriate concepts in one context, but in another context they may overgeneralize and fail to consistently apply or interpret a concept. When students are developing expertise in a new paradigm, they may be in a "knowledge in pieces" phase [33] and are more likely to overgeneralize concepts and apply principles which are inapplicable in a particular context.

### D. Inappropriate or negative transfer

Transfer of learning is defined as the application of knowledge and skills acquired in one context to another context [49]. Transfer occurs when learning in one context either enhances or undermines a related performance in another context. Negative transfer occurs when learning in one context negatively impacts performance in another context, and it commonly occurs in the early stages of learning in a new domain [50]. In introductory physics, students often have naïve notions about velocity, momentum, or work from everyday experience which can negatively transfer into their learning of classical mechanics. Similarly, concepts learned in classical mechanics, such as position and momentum, are embedded differently in the context of quantum mechanics and students may negatively transfer these concepts to quantum mechanics while developing expertise [51]. We discuss examples of these types of transfer difficulties in quantum mechanics and classical mechanics.

#### 1. Negative transfer in quantum mechanics

##### a) Difficulties with the physical, laboratory space vs. Hilbert space

The following example demonstrates negative transfer in quantum mechanics from previous courses. One common difficulty that upper-level students in quantum mechanics have is that they assume that an object with a label "$x$" is orthogonal to or cannot influence an object with a label "$y$." This difficulty is evident from responses in a multi-university study in the context of Larmor precession in which students provided responses such as "the magnetic field is in the $z$-direction so the electron is not influenced if it is initially in an eigenstate of $\hat{S}_x$" or "eigenstates of $\hat{S}_x$ are orthogonal to eigenstates of $\hat{S}_y$" [51]. In introductory physics, $x$, $y$, and $z$ are conventional labels for orthogonal components of a vector. These types of difficulties indicate that upper-level students in quantum mechanics courses negatively transferred the knowledge acquired in previous courses and were confused about the significance of the labels $x$, $y$, and $z$ to denote orthogonal spin states in a Hilbert space in quantum mechanics. In particular, students who claimed that the magnetic field is orthogonal to the eigenstate of a spin component did not realize that the magnetic field is a vector in the three-dimensional laboratory space but eigenstates of spin components are vectors in an abstract Hilbert space in which the state of the system lies.

##### b) Difficulties with successive measurements of an observable whose corresponding operator has a continuous vs. discrete eigenvalue spectrum

Another difficulty involving negative transfer in upper-level quantum mechanics courses involves measurement of an observable whose corresponding operator has continuous vs. discrete eigenvalues. Students incorrectly claimed that successive measurements of observables whose corresponding operators have a continuous eigenvalue spectrum produce somewhat deterministic outcomes whereas successive measurements of observables whose corresponding operators have a discrete eigenvalue spectrum produce very different outcomes [51]. For example, in an individual interview, one student stated: "If an observable has a continuous spectrum … the next measurement won't be very different from the first one. But if the spectrum is discrete then you will get very different outcomes." When asked to elaborate, the student added, "For example, imagine measuring the position of an electron. It is a continuous function so that time dependence is gentle and after a few seconds you can only go from A to its neighboring point (pointing to an $x$ vs. $t$ graph that he sketched on the paper). You cannot go from this without going through this intermediate space. Think of discrete variables like spin … they can give you very different values in a short time because the system must flip from up to down. I find it strange that such large changes can happen almost instantaneously. But that's what quantum mechanics predicts" [51].

##### c) Difficulties with the uncertainty principle

Another example of negative transfer from classical mechanics to quantum mechanics involves the uncertainty principle. The following question was administered to students after traditional instruction on the uncertainty principle [10]:

*Consider the following statement: "The uncertainty principle makes sense. When the particle is moving fast, the position measurement has uncertainty because you cannot determine the particle's position precisely. It's a blur and that's exactly what we learn in quantum mechanics. If the particle has a large speed, the position measurement cannot be very precise." Explain why you agree or disagree with the statement.*

The statement is incorrect because it is not the particle's speed, but rather, the uncertainty in the particle's speed that is related to the uncertainty in position. Fifty-eight percent of the students provided incorrect responses. One student stated: "I agree because when a particle has a high velocity, it is difficult to measure the position accurately." Another student agreed with the statement and provided the following reasoning: "When a particle is moving fast, we cannot determine its position exactly—it resembles a wave—at fast speed, its momentum can be better determined." Further discussions with these students indicate that students may have negatively transferred ideas from classical mechanics into quantum mechanics.

### d) Difficulties with quantum tunneling

Students also have difficulty with the concept of quantum tunneling. Research has shown that students often transfer classical reasoning when thinking about quantum tunneling [52]. Many students state that a particle "loses energy" when it tunnels through a rectangular potential barrier. This reasoning is incorrect because the particle does not lose energy when tunneling through the barrier, although the wave function of the particle inside the potential barrier is described by exponential decay. During interviews with students, common responses regarding tunneling involve statements such as: "the particle collides with and loses energy in the barrier" and "it requires energy to go through the barrier" [52]. These types of responses indicate that many students attempt to apply classical concepts to quantum mechanical situations.

## 2. Negative transfer in classical mechanics

### a) Difficulties with the definition of work

Physics education research is filled with investigations of alternative conceptions of students due to negative transfer of knowledge (see, e.g., Refs. [22, 23, 34]). For example, according to the definition of work in physics, there is no work done by a force if there is no component of force along an object's direction of motion. Introductory physics students have a naïve mental model that non-zero work must be done by the gravitational force if a person holds an object in his/her hand at rest because the person holding it gets tired. They transfer this naïve mental model into their learning of Newtonian concepts, resulting in learning difficulties.

### b) Difficulties with the net force on an object in circular motion

When an object is moving in a circle at a constant speed, it has a net force acting on it that gives rise to the centripetal acceleration. However, many students in introductory physics courses claim that there is no net force on an object in uniform circular motion because they overgeneralize concepts and associate "constant speed with no net force" even if the direction of the velocity is changing in uniform circular motion.

## 3. Possible causes for negative transfer in quantum mechanics and introductory classical mechanics

All of the above examples demonstrate that students are applying their knowledge of an older paradigm in the new paradigm. Introductory students inappropriately transfer naïve notions about motion when learning classical mechanics, and upper-level students inappropriately transfer concepts learned in classical physics to quantum mechanics. Students are attempting to fit their prior conceptions in the new paradigm's "box." Furthermore, students' prior preparation, goals, and motivation impacts their reasoning and self-regulatory skills, which can affect the extent to which they transfer knowledge appropriately. Students with limited problem-solving, reasoning, and self-regulatory skills may have difficulty in determining how a particular concept can be applied in various contexts, or they may lack the motivation to do so because they have differing goals for the course.

## E. Lack of transfer

### 1. Lack of transfer in quantum mechanics

Students often have difficulty transferring knowledge from one context to another. For example, students who had previously learned about the time dependence of a non-stationary state wave function in the context of problems involving an infinite square well were asked to find the wave function after a time $t$, given that the initial wave function was a non-stationary state wave function for a harmonic oscillator potential energy. Many students were unable to solve the problem correctly and complained that the time dependence of wave functions was only discussed in class in the context of an infinite square well so they were not sure how to solve such problems in the context of a harmonic oscillator potential energy [53].

### 2. Lack of transfer in introductory classical mechanics

In one study, 81 students in an introductory mechanics class were given a problem involving a ballerina that is commonly used by instructors in the context of angular momentum conservation [54]. In a multiple-choice format, students were asked what happens to the ballerina's angular momentum and her angular speed when she pulls her arms close to herself. In response to this question, 53% of the students provided the correct answer. An equivalent group of 65 students was given an isomorphic problem (i.e., a problem that can be mapped onto another problem in terms of the physics principle involved, although the contexts are different) in which a spinning neutron star is collapsing under its gravitational force and asked to determine what happens to the angular momentum and angular speed of the neutron star. Only 23% of the students provided the correct response. Many students did not discern the relevance of the ballerina problem that they had learned in class to the neutron star problem.

### 3. Possible causes for lack of transfer in quantum mechanics and introductory classical mechanics

The above examples indicate that students in introductory physics and quantum mechanics are often unable to transfer knowledge from one context to another. They are unable to see the deeper, underlying principles used to solve the problems. This may be due, in part, to the lack of preparation in students' problem-solving, reasoning, and metacognitive skills. Students may not have robust abstract reasoning skills to identify how different situations are isomorphic. Furthermore, students developing expertise in a new paradigm are in a "knowledge in pieces" phase [33], and so they may be unable to determine the relationship between different types of isomorphic problems.

## F. "Gut-feeling" responses inconsistent with the laws of physics

### 1. "Gut-feeling" responses in quantum mechanics

A common difficulty in quantum mechanics (analogous to introductory physics) is manifested by the fact that many students resist writing down quantum mechanical principles explicitly, and instead, answer questions based on their "gut-feeling." For example, in a multi-university study, 48% of students incorrectly claimed that $\hat{H}\Psi(x) = E\Psi(x)$ is the most fundamental equation of quantum mechanics and 39% incorrectly claimed that it is true for all possible wave functions [21]. In individual interviews, students were explicitly asked whether this equation is true for a linear superposition of the ground and first excited states of a one-dimensional infinite square well. Many students incorrectly claimed that it is indeed true in that case primarily because they incorrectly thought that the time-independent Schrödinger equation is the most fundamental equation of quantum mechanics. When these students were asked to explicitly show that this equation is true in this given context, most of them verbally argued without writing down any equations that $\hat{H}\Psi_1(x) = E_1\Psi_1(x)$ and $\hat{H}\Psi_2(x) = E_2\Psi_2(x)$ implied that their addition will give $\hat{H}\Psi(x) = E\Psi(x)$. Even when students were told that $\hat{H}\Psi(x) = E\Psi(x)$ is not obtained by summing the two individual equations, many had difficulty believing the interviewer until they explicitly wrote these equations on paper after additional encouragement to do so from the interviewer (and checked that since $E_1$ and $E_2$ are not equal, $\hat{H}\Psi(x) \neq E\Psi(x)$ for a linear superposition of energy eigenstates).

### 2. "Gut-feeling" responses in introductory classical mechanics

Students in introductory classical mechanics often use their "gut-feeling" to solve qualitative problems instead of explicitly writing down a physics principle and checking its applicability in a particular situation. If the same question is asked in a quantitative format, students are more likely to think about the applicable laws of physics. For example, 138 introductory students were asked to find a mathematical expression for the magnitude of the momentum of a boat that started from rest and had a constant horizontal force of magnitude *F* acting on it for a time *t* (and in which that force was used to tow the boat a distance *d*) [54]. Students were asked the following quantitative question:

*A tugboat pulls a ship of mass M into the harbor with a constant tension force F in the horizontal tow cable. Both the tugboat and the ship start from rest. After the ship has been towed a distance d in time t, the magnitude of its momentum will be*

(a) $Fd$
(b) $(1/2)(F/M)t^2$
(c) $(F/M)t^2/d$
(d) $(1/2)(F/M)dt^2$
(e) $Ft$

Another equivalent group of 215 introductory students was asked a similar but conceptual question in which two boats started from rest and had the same constant net horizontal force acting on each for the same period of time [54]. They were asked the following conceptual question:

*Two identical tugboats pull other ships starting from rest. The Queen Mary is a much more massive ship than the Minnow. Both tugboats pull with the same horizontal force. Neglect other forces. After both tugboats have been pulling for the same amount of time, which one of the following is true about the Queen Mary and the Minnow?*
*(a) The Queen Mary will have a greater magnitude of momentum.*
*(b) The Minnow will have a greater magnitude of momentum.*
*(c) Both ships will have the same magnitude of momentum.*
*(d) Both ships will have the same kinetic energy.*
*(e) The Queen Mary will have a greater kinetic energy*

Many introductory students used their incorrect "gut-feeling" rather than applying the appropriate physics principle (impulse-momentum theorem) to answer the conceptual question. The percentage of students providing the correct response for the qualitative question was roughly half of the percentage of students who correctly answered the quantitative problem. When a third equivalent group of 289 students (different from the first two groups) was given both questions with the quantitative question first and the qualitative question second, they performed equally well on both. Interviews suggest that introductory students who solved the quantitative problem took advantage of their expression (*Ft*) to answer the qualitative question. However, during interviews, introductory students who were only given a qualitative question wanted to use their "gut-feeling" and were very reluctant to convert the problem into a quantitative expression in order to solve it [54].

### 3. Possible causes for incorrect "gut-feeling" responses in quantum mechanics and introductory classical mechanics

The reluctance of introductory students to use their cognitive resources for quantitative analysis of qualitative problems is similar to the reluctance of advanced students in quantum mechanics to verify the validity of $\hat{H}\Psi(x) = E\Psi(x)$ explicitly by writing it down in the given situation. One possible explanation for students using their "gut-feeling" is that many students lack robust problem-solving, reasoning, and self-regulatory skills. Furthermore, students are still developing expertise in a significantly new paradigm (the classical mechanics paradigm is different from students' naïve mental models and the quantum mechanics paradigm is different from the classical mechanics paradigm). Consequently, writing down each step explicitly and converting a conceptual question to a quantitative question in order to solve it are cognitively demanding tasks and may cause cognitive overload [55]. This may lead some students to solve problems based on their "gut-feeling" rather than by engaging in the cognitively demanding task of generating systematic solutions using physics principles.

### G. Difficulties in solving multi-part problems
#### 1. Difficulties in solving multi-part problems in quantum mechanics

The following example demonstrates student difficulties with solving multi-part problems in quantum mechanics. Upper-level students were first given an initial wave function of a particle in an infinite square well which was not a stationary state. They were then told that a measurement of position was performed. Students were asked to describe the quantum wave function of the particle a long time after the position measurement. According to the Copenhagen interpretation of quantum mechanics (the most widely-held view of the nature of measurement in quantum mechanics and the most commonly taught interpretation of quantum mechanics), a particle in generic superposition of states is forced into a single state by the act of measurement. After a position measurement, the particle will become localized in space and the corresponding position space wave function will collapse into a delta function centered about the measured position. With time, the highly-peaked wave function will evolve according to the Hamiltonian, but the wave function is neither "stuck" in the collapsed state nor will it go back to the original state before the position measurement. However, many students who had already taken an upper-level quantum mechanics course claimed that a long time after a position measurement, the wave function of the system will go back to the state before the measurement was performed [56]. Other students who provided incorrect responses often claimed that the wave function "gets stuck" in the collapsed state after a position measurement [56]. In individual interviews, these students were explicitly told that their initial responses were not correct and that they should think about what quantum mechanics predicts about the wave function a long time after the position measurement. Then, students who initially claimed that the wave function reverts to the original wave function a long time after the position measurement typically changed their response, saying that the wave function gets stuck in the collapsed state. The students who initially claimed that the wave function gets stuck also typically changed their response, saying that the wave function reverts to the original wave function. When the students were told that neither of the possibilities are correct and that they should think about what quantum mechanics actually predicts, some of them explicitly asked the interviewer how any other possibility exists for this situation because these are the only two possibilities they could generate. The fact that the delta function will start evolving according to the time-dependent Schrödinger Equation (TDSE) based upon the Hamiltonian of the system was something these advanced students were unable to take into account.

### 2. Difficulties in solving multi-part problems in introductory classical mechanics

A similar difficulty in introductory physics is observed with a three-part problem involving a ballistic pendulum in which a piece of putty is raised to a certain height and released. It then collides with another piece of putty, the two pieces of putty stick together, and then the merged pieces of putty rise together [57]. In a multi-university study, students were asked for the final height of the merged pieces of putty in terms of the initial height of one of the pieces of putty. Even after instruction, only 27% of the introductory students noted that both conservation of energy and conservation of momentum should be used to answer this question. A majority of students incorrectly claimed that only one of these principles is sufficient to find the final height of the merged putties in terms of the initial height because they either focused on the change in height of the putty or the collision [57].

### 3. Possible causes for difficulties in solving multi-part problems in quantum mechanics and introductory classical mechanics

The difficulties in solving multi-part problems may be caused by students' inadequate problem-solving, reasoning, and self-regulatory skills—many students may not have a sufficient skill set to break the problem into sub-problems and coordinate different principles and concepts in which the outcomes of the different sub-problems are coupled to each other. Furthermore, students who are still developing expertise in a new paradigm may only focus on some parts of the problem while solving a complex problem. Since students in each sub-domain of physics are still developing expertise and their knowledge is in pieces [33], it is often difficult for them to solve complex, multi-part problems.

## H. Difficulties related to students' epistemological views

According to Hammer [35], a student's epistemology regarding physics includes three components: 1) beliefs about the structure of physics knowledge as a collection of isolated pieces or a single coherent system; 2) beliefs about the content of physics knowledge as formulas or concepts that underlie the formulas; and 3) beliefs about learning physics, whether it entails receiving information or actively reconstructing one's understanding. Students' epistemologies can impact whether they engage in self-regulation, sense-making, and building a robust, conceptual knowledge structure. Similar to students in introductory physics [35], students' inadequate preparation, unclear goals, and insufficient motivation and the fact that the paradigm of quantum mechanics is significantly different from classical mechanics can influence students' epistemological views of quantum mechanics, e.g., whether they check for mathematical consistency and strive to develop a good knowledge structure instead of memorizing algorithms. Below, we discuss examples of students' epistemological views on reconciling physical models with their mental models, checking for consistency in their answers, reliance on memorizing algorithms, ambiguous or careless language, and the learning process in quantum mechanics and introductory mechanics. Finally, we discuss some possible causes which may explain why students exhibit difficulties in developing expert-like epistemological views.

1. **Difficulties in reconciling physical models with one's own mental model**
   a) Difficulties in reconciling physical models with one's own mental model in quantum mechanics

Students often have difficulty describing the measurement process in quantum mechanics. For example, in a multi-university study, the following problem was administered to 202 students [21]:

*The wave function of an electron in a one-dimensional infinite square well of width a at time t=0 is given by* $\Psi(x, 0) = \sqrt{2/7}\, \phi_1(x) + \sqrt{5/7}\, \phi_2(x)$, *where* $\phi_1(x)$ *and* $\phi_2(x)$ *are the ground state and first excited stationary state of the system.* ($\phi_n(x) = \sqrt{2/a}\, sin(n\pi x/a)$, $E_n = n^2\pi^2\hbar^2/2ma^2$, *where* $n = 1,2,3 ...$).
a) *You measure the energy and the measurement yields* $4\pi^2\hbar^2/2ma^2$. *Write an expression for the wave function right after the measurement.*
b) *Immediately after this energy measurement, you measure the position of the electron. Qualitatively describe the possible values of position you can measure and the probability of measuring them.*

In response to part a), some students claimed that the system should remain in the original state which is a linear superposition of the ground and first excited states after the energy measurement. One student stated: "… the collapse of the wave function is temporary … . Something has to happen to the wave function for you to be able to measure energy or position, but after the measurement the wave function must go back to what it *actually* (student's emphasis) is supposed to be." Students with this type of reasoning often felt that the collapse of the wave function during a measurement is a "trick" used in the Copenhagen interpretation to find the possible outcomes and their probabilities but the wave function must revert back to what it *actually* represents (which is the wave function right before the measurement). Some students claimed that their instructor had explicitly mentioned that the collapse of the wave function is not real but just a "trick." They incorrectly interpreted it to mean that the collapse does not really change the wave function. In response to part b), some students incorrectly noted that because the energy is well defined immediately after the measurement of energy, the uncertainty in position must be infinite according to the generalized uncertainty principle. When a student with this type of response was asked to plot the wave function after the energy measurement, the student was able to do that correctly but he still continued to claim that the uncertainty in position must be infinite in this state. When the student was explicitly asked about how one would calculate the uncertainty in position, he was unable to articulate it correctly although he noted that it has something to do with how accurately you can measure the position. He admitted that he had difficulty forming good pictures in his mind about quantum measurement. In response to part b), one student argued that it may not be possible to measure the position after measuring the energy, stating: "Can you even do that? Doesn't making a measurement change the system in a manner that makes another measurement invalid?" This student was struggling with what the incompatibility of observables means and whether incompatibility implies that it is impossible to measure one incompatible observable after another (which seems absurd from an experimental point of view). These types of statements shed light on students' epistemological views about quantum theory. Advanced students learning quantum mechanics struggle to come up with good mental models of quantum measurements and some of them may have difficulty reconciling their own mental models with the appropriate physical model.

   b) Difficulties in reconciling physical models with one's own mental model in classical mechanics
      (1) Difficulties in interpreting situations for which mechanical energy of a system is not conserved

Similar to upper-level students, students in introductory physics often develop their own mental models of classical concepts. They may have difficulty reconciling their mental model with the appropriate physical model in a given context. In a survey on energy and momentum [57], the following situation was posed to students:

*Three bicycles approach a hill as described below:*
*(1) Cyclist 1 stops pedaling at the bottom of the hill, and her bicycle coasts up the hill.*
*(2) Cyclist 2 pedals so that her bicycle goes up the hill at a constant speed.*
*(3) Cyclist 3 pedals harder, so that her bicycle accelerates up the hill.*
*Ignoring the retarding effects of friction, select all the cases in which the <u>total mechanical energy</u> of the cyclist and bicycle is conserved.*
*(a) (1) only*

*(b) (2) only*
*(c) (1) and (2) only*
*(d) (2) and (3) only*
*(e) (1), (2), and (3).*

In this scenario, going up a hill at a constant speed implies that mechanical energy is not conserved because mechanical energy must be put into the system by a non-conservative force in order to make the bicycle go up at a constant speed. One interviewed student who chose the incorrect option (e) explained: "if you ignore the retarding effects of friction ... mechanical energy will be conserved no matter what." Other interviewed students who chose option (e) also suggested that the retarding effect of friction was the only force that could change the mechanical energy of the system. Although some students may have chosen (b) because they could not distinguish between the kinetic and mechanical energies, the following interview excerpt shows why that option was chosen by a student despite the knowledge that kinetic and mechanical energies are different: "if she goes up at constant speed then kinetic energy does not change ... that means potential energy does not change so the mechanical energy is conserved ... mechanical energy is kinetic plus potential." When asked to explain what the potential energy is, the student continued: "isn't it *mgh*?" When asked to explain why it is not changing, the student first paused and then added: "*h* is the height ... I guess *h* does change if she goes up the hill ... maybe that means that potential energy changes. I am confused ... I thought that if the kinetic energy does not change, then potential energy cannot change ... aren't the two supposed to compensate each other ... is it a realistic situation that she bikes up the hill at constant speed or is it just an ideal case?" The student was convinced that the mechanical energy is conserved because he was asked to ignore the retarding effects of friction (which he thought was the only non-conservative force that can do work on the system and change mechanical energy). Thus, he used his mental model that mechanical energy was conserved in that case to claim that both the kinetic and potential energies must remain unchanged (even though potential energy must be changing as the bike goes up the hill). When he confronted the fact that the potential energy is changing, he failed to reason that the mechanical energy must be changing if the kinetic energy is constant. Instead, he questioned whether it is realistic to bike up the hill at a constant speed and suggested that this is only possible in the idealized physics world. Although he ignored the work done by the non-conservative force applied on the pedal to keep the speed constant, his statements shed light on students' epistemological beliefs about how much one can trust physics to explain everyday phenomena and his difficulty in reconciling his mental model with the physics model.

### (2) Difficulties with normal force

The following example involving the normal force on an inclined plane demonstrates how introductory students have difficulties in reconciling their own mental models with appropriate physics models. When learning about normal force, many students create a mental model in which the force due to gravity is always antiparallel to the normal force. This model is only appropriate for objects on a horizontal surface. However, in the context of an inclined plane, many students incorrectly claim that the normal force is not perpendicular to the inclined plane, but rather, antiparallel to the force due to gravity. When questioned about their answer, students often state that this is what their instructor told them. Students are interpreting what their instructor taught them to conform to their mental model. Similar difficulties are displayed when children learn about the shape of Earth. Since children often have the mental model that Earth is flat, when they are told that Earth is round, they often claim that the earth is round like a flat pancake or that it is shaped like a hemisphere and humans live on the flat side [58]. In these cases, students are coming up with mental models that may take into account some elements of what they are taught but are modified to make them consistent with their own world view.

## 2. Difficulties involving overlooking consistency

### a) Difficulties involving overlooking consistency in quantum mechanics

Another type of difficulty in reasoning and self-monitoring is displayed when students explicitly violate mathematical rules of linear algebra in the context of quantum mechanics. For example, students were asked the following question about a quantum mechanical operator $\hat{Q}$ acting on a generic state $|\Psi\rangle$:

*Consider the following conversation between Andy and Caroline about the measurement of an observable Q for a system in a state $|\Psi\rangle$ which is not an eigenstate of $\hat{Q}$:*

**Andy**: *When an operator $\hat{Q}$ corresponding to a physical observable Q acts on the state $|\Psi\rangle$, it corresponds to a measurement of that observable. Therefore, $\hat{Q}|\Psi\rangle = q_n|\Psi\rangle$ where $q_n$ is the observed value.*
**Caroline**: *No. The measurement collapses the state so $\hat{Q}|\Psi\rangle = q_n|\Psi_n\rangle$ where $|\Psi_n\rangle$ on the right hand side of the equation is an eigenstate of $\hat{Q}$ with eigenvalue $q_n$.*
*With whom do you agree?*
  a) *Agree with Caroline only*
  b) *Agree with Andy only*
  c) *Agree with neither*
  d) *Agree with both*
  e) *The answer depends on the observable Q.*

Fifty-two percent of the students claimed that $\hat{Q}|\Psi\rangle = q_n|\Psi_n\rangle$, $\hat{Q}|\Psi\rangle = q_n|\Psi\rangle$, or both equations are correct [59]. Actually, neither of the above equations is correct because they both violate basic rules of linear algebra. In one-on-one interviews, students were so focused on thinking about how a single equation should describe the measurement process and the collapse of the wave function that none of them felt the need to verify that the above equations are both incorrect in terms of linear algebra. Upper-level students are unlikely to make such mistakes if this question is asked in a linear algebra course without the quantum mechanics context. However, in the context involving quantum measurement, their incorrect conception that an operator corresponding to an observable acting on a quantum state corresponds to the measurement of the observable was so strong that they did not consistently apply tenets of linear algebra. When the interviewed students were explicitly asked how the right hand side (RHS) of an equation can change when the left hand side (LHS) remains the same, many students appeared not to be concerned about such an anomalous situation in linear algebra where, depending upon the context, the same LHS yields a different RHS. Students were often very focused on the context. They were convinced that the collapse of a wave function upon the measurement of an observable in quantum mechanics must be represented by an equation and Caroline's equation must correspond to the equation after the collapse of the wave function has occurred. They often reiterated that such changes occur only to the RHS (and the LHS is the same for both Andy and Caroline) because the RHS corresponds to the "output" and the LHS corresponds to the "input." According to their reasoning, it is only the output that is affected by the measurement process (and not the input) so the LHS for Andy and Caroline are the same. When students were asked to explicitly choose the observable to be energy so that the operator is the Hamiltonian operator, their qualitative responses were unchanged even in that concrete case. Some interviewed students were explicitly asked to explain how, on the RHS, a linear combination of the eigenfunctions of the operator that Andy proposes can be the same as only one term in the sum proposed by Caroline in her equation. These students often explained their reasoning by claiming that an operator acting on the wave function corresponds to the measurement of the observable as Caroline proposes. They incorrectly added that Andy's equation is true only before the measurement of the observable has actually taken place and Caroline's statement is true right after the measurement of the observable has taken place and has therefore led to the collapse of the wave function. Many students explicitly stated that at the instant the measurement takes place both Caroline and Andy are correct because the wave function undergoes an instantaneous collapse and the RHS of the equation changes. As noted, in the interviews, when students were explicitly reminded that the equation they thought was correct violated linear algebra and the RHS of an equation cannot change when the left hand side remains the same, some students became worried. However, some students noted that they were unsure about the rules of quantum mechanics and that they were not sure whether quantum mechanics not only violates the principles of Newtonian mechanics but also violates the rules of linear algebra.

The above example shows how difficult the measurement postulate (based upon the Copenhagen interpretation) is from an epistemological point of view and how students have built a locally coherent knowledge structure (inconsistent with the quantum postulate) to represent the measurement process with equations. It is also interesting to note that since students were often convinced about the physical process of the wave function collapse being represented by the equations that Andy and Caroline wrote, they overlooked the linear algebra involved and did not question the anomaly regarding the same LHS yielding different RHS. Unproductive epistemological views, e.g., "quantum mechanics is not supposed to make sense" or "perhaps linear algebra is not necessarily supposed to work as expected in quantum mechanics" may lead to serious difficulties in reasoning and can impede sense-making.

b) Difficulties involving overlooking consistency in introductory classical mechanics

Similar overlooking of mathematical or other types of consistency, especially due to strong alternative conceptions, is common in introductory physics. For example, in one investigation, introductory students were given two isomorphic problems involving Newton's second law in an equilibrium situation on an inclined plane [54]. The two questions are shown below:

*A car which weighs 15,000 N is at rest on a frictionless 30° incline. The car is held in place by a light strong cable parallel to the incline. Find the magnitude of tension force T in the cable.*
*a) 7,500 N*
*b) 10,400 N*
*c) 11,700 N*
*d) 13,000 N*
*e) 15,000 N*

*A car which weights 15,000 N is at rest on a 30° incline. The coefficient of static friction between the car's tires and the road is 0.90, and the coefficient of kinetic friction is 0.80. Find the magnitude of the frictional force on the car.*
*a) 7,500 N*
*b) 10,400 N*
*c) 11,700 N*
*d) 13,000 N*
*e) 15,000 N*

The second problem elicits a strong incorrect conception that static frictional force is always at its maximum value. Many introductory students ignored the similarity between the adjacent problems (including the fact that the free-body diagrams provided were identical except that the tension force in one problem was replaced by the frictional force in the other problem, which would logically imply that the desired quantities, tension and friction, had the same magnitude). While 72% of the students answered the tension problem correctly, only 28% of the students provided the correct response to the friction problem. A majority did not recognize the mathematical consistency between the isomorphic problems—despite doing the tension problem correctly—and launched into a calculation of maximum static friction although it was not at the maximum value in the problem. Asking students to explicitly focus on the free body diagrams for each isomorphic problem did not help [54]. Even when students' attention was explicitly brought to the fact that the free body diagrams were similar except that the tension force was replaced by the friction force, students continued to stick with their initial answer, claiming that one does not need to use the free body diagram for the friction force for which there is a formula but one must use the free body diagram for the tension force for which there is no formula. These types of epistemological views about learning introductory physics can lead to a lack of incentive to look for coherence and a unified nature of physics knowledge and can impact how much effort students make to build a robust knowledge structure.

### 3. Difficulties due to reliance on memorized algorithms

a) Difficulties due to reliance on memorized algorithms in quantum mechanics

Many upper-level students in quantum mechanics use memorization tactics over conceptual understanding—preferring to "plug and chug" without understanding the underlying concepts. For example, in a multi-university survey, the following problems were administered to 202 students [21]:

*The wave function of an electron in a one-dimensional infinite square well of width a at time t=0 is given by $\Psi(x,0) = \sqrt{2/7}\,\phi_1(x) + \sqrt{5/7}\,\phi_2(x)$, where $\phi_1(x)$ and $\phi_2(x)$ are the ground state and first excited stationary states of the system. ($\phi_n(x) = \sqrt{2/a}\sin(n\pi x/a)$, $E_n = n^2\pi^2\hbar^2/2ma^2$, where $n = 1,2,3\ldots$). Answer the following questions about this system:*
*a) Write down the wave function $\Psi(x,t)$ at time t in terms of $\phi_1(x)$ and $\phi_2(x)$.*
*b) You measure the energy of an electron at time t=0. Write down the possible values of the energy and the probability of measuring each.*

c) *Calculate the expectation value of the energy in the state $\Psi(x,t)$ above.*

For part c), the expectation value of energy is time independent because the Hamiltonian does not depend on time. The expectation value of energy in this state is $\langle E \rangle = 2/7\, E_1 + 5/7\, E_2$. Only 39% of the students provided the correct response to part c) although 67% of students answered part b) correctly. Many students did not discern the relevance of part b) for part c) and did not exploit what they found for part b) in answering part c). Consequently, many of the students worked out the expectation value of energy from scratch by explicitly writing $\langle E \rangle = \int \Psi^*(x,t)\hat{H}\Psi(x,t)dx$. Then, they wrote the wave function as a linear superposition of the ground state and first excited state and were able to show that the Hamiltonian operator $\hat{H}$ acting on the stationary states will give the corresponding energy and the same state back. They further demonstrated that the time-dependent phase factors for the two terms that survive will vanish due to complex conjugation. However, many students who tried to solve the problem using the memorized algorithm for calculating expectation value got lost along the way. They often forgot to take the complex conjugate of the wave function, use orthogonality of stationary states, or did not realize the proper limits of the integral. This example sheds light on the epistemology of students in upper-level quantum mechanics and suggests that sometimes even they rely on memorized knowledge and employ complicated algorithmic approaches instead of focusing on the significantly simpler approach that exploits the underlying quantum concepts to solve problems.

b) Difficulties due to reliance on memorized algorithms in introductory classical mechanics

Research in introductory physics teaching and learning suggests that introductory students often use a "plug and chug" approach to problem solving [23]. Research suggests that introductory students are able to solve seemingly difficult problems because they can apply an algorithm to get the correct final answer but fail to answer simpler conceptual questions related to the same problem. Mazur illustrates this with examples and states, "it is possible for students to do well on conventional problems by memorizing algorithms without understanding the underlying physics" ([60], p. 6). In solving quantitative problems, students often look for a formula consistent with the givens and variables in the problem and proceed with an algorithmic approach without thinking about the physics principles involved. For example, a student who knows how to use the algorithm for conservation of mechanical energy can derive an expression for the speed of a person at the bottom of a slide who started at rest from the top but may be unable to answer whether the speed at the bottom of the slide depends on the mass of the person if asked as a qualitative question [57].

### 4. Difficulties due to the interpretation of ambiguous or careless language
#### a) Difficulties with ambiguous or careless language in quantum mechanics
##### (1) Difficulties with the wave-particle duality

The terminology of quantum mechanics incorporates some of the vocabulary and concepts used in classical mechanics, e.g., position, momentum, and energy, but these concepts are not utilized or interpreted in the classical way. As an example, the double-slit experiment demonstrates the wave-particle duality of single electrons. An electron passes through both slits while traveling toward a screen due to its wave-like nature. However, when the electron arrives at the detecting screen, a flash is seen at one location on the screen due to the collapse of the wave function. These types of experiments are epistemologically challenging even for advanced students. The wave-particle duality of a single electron that becomes evident at different times in the same experiment is very difficult for students to rationalize. Students may have used vocabulary such as "particle" to describe a localized entity in their classical mechanics courses. Consequently, they may find it very difficult to think of the electron as a wave in part of the experiment and as a particle in another part of the experiment (when it lands on the detecting screen and the wave function collapses). To reduce this difficulty, some researchers coined the term "wavicle" [61] for a quantum entity such as an electron. However, this terminology did not become popular.

##### (2) Difficulties with terminology involving the Mach-Zehnder Interferometer with single photons

The use of careless vocabulary by experts and novices alike is a challenge students face in learning quantum mechanics. A new paradigm may require new vocabulary to explain radically different concepts. However, the concepts in quantum mechanics are often expressed using classical terminology. For example, in quantum mechanical gedanken (thought) experiments, terminology such as "which-path" or "which-slit" information was popularized by Wheeler [62]. One experiment which often elicits careless vocabulary by instructors is the Mach-Zehnder interferometer. Similar to the double-slit experiment, the Mach-Zehnder interferometer with single photons is an experiment which has been conducted in undergraduate laboratories to illustrate fundamental principles of quantum mechanics (see Figure 4) [63]. In this experiment, a large number of single photons are emitted from the source. After propagating through beam-splitter 1 (BS1), a photon is in a superposition of the upper (U) and lower (L) path states. Beam-splitter 2 (BS2) mixes the path states of a single photon and the detectors (D1 and D2) can project both components of the photon path state, which interfere at the detectors. The single photon path states from the two paths arriving at detector 1 (D1) undergo a total phase shift of $2\pi$, arriving in phase at D1 and displaying constructive interference. If the photon source emits a large number of photons, all photons will arrive at D1. The single photon path states from the two paths arriving at detector 2 (D2) arrive out of phase, displaying destructive interference. If the source emits a large number of single photons, no photons will arrive at D2. Changing the thickness of the phase shifter will affect how many photons arrive at the detectors.

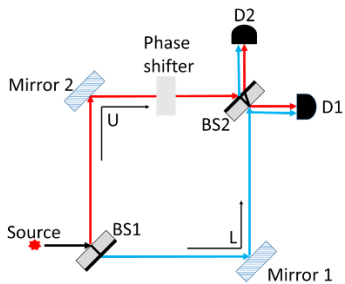

*Figure 4. Mach-Zehnder Interferometer setup with a phase shifter in the upper path*

Wheeler suggested that observing interference of a single photon with itself at D1 and D2 when a large number of single photons are emitted from the source can be interpreted in terms of not having "which-path" information about the single photon [62]. In the Mach-Zehnder interferometer experiment, following Wheeler, it is often stated that "which-path" information is unknown if the photon "took both paths" and displays interference effects at the detectors (see Figure 4). However, if beam-splitter 2 is removed after the photon has already propagated through beam-splitter 1 as in the delayed-choice experiment (see Figure 5), it is said that "which-path" information is known because the photons arriving at D1 must have propagated through the upper path only and the photons arriving at D2 must have propagated through the lower path only [62]. When discussing the delayed-choice experiment in the Mach-Zehnder interferometer, many instructors use Wheeler's terminology and state that "all photons reaching D2 took the lower path and all photons reaching D1 took the upper path." However, this type of terminology may indicate that one can retro-cause the photon to go through both paths or one path by inserting or removing beam-splitter 2 after the photon has propagated through beam-splitter 1 [64]. Students may develop unproductive epistemologies that quantum mechanics phenomena can violate causality. In the situation in which beam-splitter 2 is inserted after the photon has already propagated through beam-splitter 1, students have additional difficulties. For example, some students claim that detector 1 would register a photon 50% of the time and detector 2 would never register a photon because, although the photon arrives at detector 2, destructive interference "kills" the photon and it is lost. These types of statements shed light on students' epistemology and the challenges in the development of expertise in quantum mechanics.

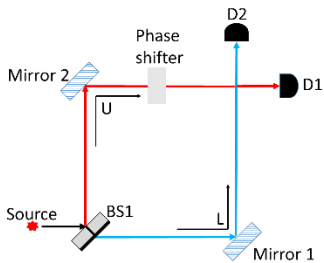

*Figure 5. Mach-Zehnder Interferometer with beam-splitter 2 removed*

To avoid this confusing terminology, we have developed a quantum interactive learning tutorial about a Mach-Zehnder interferometer with single photons [65]. In this tutorial, students learn about how a photon is in a superposition of both the upper and lower path states after propagating through beam-splitter 1. The photon is in a superposition of path states until it arrives at the detectors, regardless of whether beam-splitter 2 is inserted or removed. Once the photon is detected at D1 or D2, the superposition state collapses. "Which-path" information is "known" about the photon if D1 and D2 can only project one component of the photon path state (as opposed to the photon "taking *either* the upper or lower path"). For example, in Figure 5, "which-path" information is known for single photons arriving at the detectors because only the component of a photon state along the upper path can be projected in D1 and only the component of a photon state along the lower path can be projected in D2. On the other hand, "which-path" information is unknown about single photons arriving at the detectors in the setup shown in Figure 4 because beam-splitter 2 mixes the path states of the single photon. Thus, D1 and D2 can project both components of the photon path state and the projection of both components at each detector leads to interference.

      b) Difficulties with ambiguous or careless language in introductory physics
         (1) Difficulties with everyday terminology vs. physics terminology

In introductory physics, even before formal instruction, many students have common notions from everyday experience, e.g., a larger constant velocity implies a larger net force, momentum is equivalent to force, velocity is equivalent to speed, acceleration is equivalent to force, and work is done by a force even if there is no displacement. However, since these concepts are defined differently in physics, their incorrect notions impede their learning. If instruction is not designed appropriately to help students explicitly resolve issues involving terminology and concepts in the new paradigm, they may conclude that physics does not make sense and physics is about idealized situations that cannot be used to understand real-world phenomena. Students may try to memorize what they are taught and combine their own mental models with physics models to come up with something that is not consistent with the laws of physics as discussed in the examples earlier.

         (2) Difficulties with defining the system when angular momentum is conserved

In introductory physics, an instructor may state that angular momentum is conserved when a ballerina holding a barbell pulls her arms close to her body or extends her arms far from her body. However, instructors may not clarify for which system the angular momentum is conserved, assuming it is obvious to the students. Nevertheless, knowing what happens to the ballerina's angular speed if she drops the barbell requires an understanding of the fact that the angular momentum is conserved for the ballerina-barbell system. This type of ambiguity about the appropriate system also exists for mechanical energy conservation, linear momentum conservation, etc. In both classical mechanics and quantum mechanics, instructional design should explicitly focus on clarity of language to guide students to learn the concepts in a new paradigm.

   **5. Difficulties associated with unproductive beliefs about active engagement during the learning process**
      a) Difficulties associated with unproductive beliefs about active engagement while learning quantum mechanics
         (1) Reliance on rote learning strategies vs. active construction of a coherent knowledge structure

Interviews suggest that, even in upper-level quantum mechanics, many students do not use their mistakes as an opportunity for learning and for building a robust knowledge structure and they resort to rote learning strategies for getting through the course [19]. For example, instead of focusing on developing a robust knowledge structure of quantum mechanics, students employed test-taking strategies (which have nothing to do with developing conceptual understanding and a coherent knowledge structure) by focusing only on fragments of the material that the instructor was likely to ask on exams and skipping the material that was on the midterm examination while studying for the final exam (because they did not expect that material from the midterm exam to be repeated on the final exam) [19].

         (2) Reliance on the instructor as the authority

In addition, students in quantum mechanics courses often make statements in interviews similar to introductory physics students [35] indicating that they believe the instructor is an authority on the subject and therefore they accept what the instructor says without questioning it. These types of attitudes can lead to students not making an effort to develop a robust knowledge structure or engage in sense-making. For example, in a multi-university survey with more than 200 students, on a question regarding whether the time-independent Schrödinger equation is the most fundamental equation of quantum mechanics, 39% of students incorrectly claimed that $\hat{H}\Psi = E\Psi$ is unconditionally true whereas it is only true for stationary states for a given system. Interviews with a subset of students suggest that, typically, they felt that this is what their instructor had taught them. For example, one student who was confident that this is what his instructor had taught stated: "This is what 80 years of experiment has proven. If future experiments prove this statement wrong, then I'll update my opinion on this subject." These students incorrectly interpreted what the instructor had said. Another interviewed student who was told later in an interview situation that possible wave functions need not satisfy $\hat{H}\Psi = E\Psi$ and any normalized smooth function that satisfies the boundary conditions is a possible wave function for a system threw up his hands and argued that "if possible wave functions can be that generic, then what is the point of the Schrödinger equation?" He stated that what he was being told by the interviewer did not sound like what his instructor had taught, and he did not know what to make of it. The student did not realize that the purpose of the time-dependent Schrödinger equation, which is the most fundamental equation of quantum mechanics, is to govern the time-evolution of the wave function (it is analogous to Newton's second law in classical mechanics). Responses of this type indicate that students may take the instructor's words without questioning but internalize the instruction by adapting it to achieve consistency with their own mental models. In turn, they may not make the effort to self-regulate or build a robust knowledge structure [21].

We note that instructors may also inadvertently hinder students' sense-making in the new paradigm of quantum mechanics by echoing the statement of Richard Feynman: "I think I can safely say that nobody understands quantum mechanics" [66]. Feynman or the instructor is referring to the fact that they do not understand the origin of the postulates and interpretations, but interviews suggest that the students often misinterpret them to mean that they do not know how to "do" quantum mechanics. Interviews with individual students also suggest that they reason that if their instructor does not understand quantum mechanics formalism, it will be impossible for them to understand it. This viewpoint can hinder students' self-regulation of learning. They may not engage deeply with the basic tenets of quantum mechanics to build a coherent knowledge structure but rather assume, e.g., that quantum mechanics is so strange that it can also violate mathematical rules of linear algebra as discussed earlier. To counteract this viewpoint, students should be made aware of the distinction between understanding the "origin" of the postulates and interpretations of quantum mechanics vs. "doing" quantum mechanics. While there are many interpretations of quantum mechanics and the underlying reasons for why the postulates of quantum mechanics work are difficult to understand (even within the Copenhagen interpretation that is taught to students), learning and applying the Copenhagen interpretation can allow students to relatively easily calculate what is desired. While it is true that many quantum phenomena are not yet fully understood (e.g., the mechanisms for exotic behavior in some highly correlated electron systems such as high temperature superconductors), quantum formalism has been highly successful in explaining and predicting outcomes of experiments. Similarly, some instructors use phrases such as "the collapse of the wave function is just a trick," but students misinterpret it to imply that the collapse does not change the state of the system even if the system was not in an eigenstate of the operator corresponding to the observable measured (as in the example discussed earlier). Indeed, misinterpretations of this type can have detrimental effects on students' epistemology and ultimately their learning.

  b) Difficulties associated with unproductive beliefs about active engagement while learning introductory physics

Similar to upper-level students' epistemological views about learning quantum mechanics, research also suggests that many introductory physics students believe that physics is simply a collection of facts and formulas and that the teacher is the authority on the subject [35, 67]. Thus, they take meticulous notes and memorize knowledge imparted rather than engaging in sense-making [35, 67]. These beliefs can hinder students' self-regulation and the building of a robust knowledge structure. Similarly, Schoenfeld emphasizes that when students are taught basic mathematics in the traditional method, they may come to believe that school mathematics consists of memorizing formal procedures taught by instructors that are completely divorced from real life [68].

  6. Possible causes for difficulties in developing expert-like epistemological views

The above examples and discussions suggest that students in quantum mechanics and introductory physics often display unproductive epistemologies that can hinder learning. The paradigm shift may partly cause difficulties and make it difficult for them to reconcile their mental models with correct physical models, make them overlook inconsistency, and make it difficult for them to clarify for themselves the ambiguous or careless language used. It can negatively impact students' views of the structure of physics knowledge and lead them to think that it is a collection of isolated pieces of information (as opposed to a single coherent system) and that the content of physics knowledge is a collection of formulas (as opposed to concepts that underlie the formulas) that comes from an authority. In addition, students' prior preparation, goals, and motivation can affect the extent to which they hold productive beliefs about active participation in the learning process, impacting their perception of how to learn physics (receiving information vs. active processing). The paradigm shift coupled with students' unclear goals, insufficient motivation, and inadequate prior preparation can greatly influence the development of expert-like epistemological views in both introductory physics and quantum mechanics.

## IV. DISCUSSION AND IMPLICATIONS OF THE FRAMEWORK FOR LEARNING QUANTUM MECHANICS

It is widely assumed that a majority of upper-level physics students have not only learned significantly more physics content, but have also developed significantly better reasoning, problem-solving, and self-regulatory skills than introductory physics students. However, expertise is domain-specific—it is unclear how readily skills transfer across domains [36-38]. Classical mechanics and quantum mechanics are two significantly different paradigms. Learning quantum mechanics can be challenging even for advanced students who have developed a good knowledge structure of classical mechanics. These challenges are similar to the challenges faced by students in introductory mechanics who are transitioning from their naïve views about force and motion to those consistent with Newtonian physics [20, 21].

As discussed earlier, many physics education researchers, e.g., McDermott, Halloun and Hestenes, Clement, etc. [22, 23, 26, 34] have emphasized that students in introductory mechanics courses often struggle because they have inadequate prior preparation and diverse goals and motivations for excelling in a course in addition to the fact that the paradigm of classical mechanics is very different from the common sense conceptions students develop trying to rationalize their everyday experiences. Although these researchers may not have explicitly attributed these reasons for introductory student difficulties to a framework, they essentially describe a framework for why many introductory physics students struggle in these courses.

In this paper, we described a framework that posits that the patterns of difficulties that students face in developing expertise in quantum mechanics are analogous to what many students face in learning introductory physics. The framework incorporates the facts that many students have inadequate preparation, unclear goals, and insufficient motivation and that the paradigms of classical mechanics and quantum mechanics are significantly different. In particular, students in both introductory classical mechanics and upper-level quantum mechanics have varying goals, motivations, and preparation including a range in the proficiency of their problem-solving, reasoning, mathematical, and self-regulatory skills. In addition, students in both introductory mechanics and quantum mechanics encounter a paradigm shift in which they must assimilate and accommodate radically different concepts from what they are used to. Because of these similarities, the patterns of student difficulties in quantum mechanics are analogous to those of introductory students learning classical mechanics. The framework helps explain analogous patterns of various types of difficulties, e.g., lack of a robust knowledge structure and effective problem-solving skills, failure to transfer knowledge, lack of self-regulation, cognitive overload, and unproductive epistemological views. This framework can be used to help instructors further contemplate possible patterns of student reasoning and metacognitive difficulties in learning quantum mechanics. It can also enable physics education researchers and curriculum developers to leverage the extensive literature for introductory physics education research and adapt promising approaches to help guide the design of effective teaching and learning strategies for quantum mechanics.

### A. Development of research-based curricula and pedagogies for quantum mechanics

Research in introductory physics suggests that in order to help all students with diverse goals and preparation build a robust knowledge structure of introductory mechanics, appropriately connect mathematics and physics, and learn to apply physics principles in diverse situations to explain and predict phenomena, instructional design should conform to the field tested cognitive apprenticeship model [69]. Our framework suggests that a similar model may be useful for helping students develop a functional understanding of quantum mechanics. The cognitive apprenticeship model of learning involves three major components: "modeling," "coaching and scaffolding," and "weaning." This approach has also been found to be effective in helping students learn effective problem-solving heuristics and developing their reasoning and metacognitive skills. In this approach, "modeling" means that the instructor demonstrates and exemplifies the skills that students should learn. "Coaching and scaffolding" refer to providing students suitable practice, guidance, and feedback so that they learn the skills necessary for good performance. "Weaning" means gradually fading the support and feedback with a focus on helping students develop self-reliance.

In many traditionally taught, "lecture-only" physics classes at all levels, instructors model criteria of good performance. Modeling is often done implicitly in lectures, which is not very effective. As adaptive experts [40], instructors are unaware of some of the cognitive processes they engage in and do not model these explicitly for the students. However, what is truly lacking in the traditional instructional approach is coaching and scaffolding. In that sense, the traditional model of teaching physics is akin to asking students to watch the instructor or the TA play piano (solve physics problems for them) and then telling them to practice playing piano on their own (solve physics problems in homework) [70]. Based upon the framework, students with a wide variety of goals and backgrounds in both introductory mechanics and quantum mechanics may struggle to develop a functional understanding in a novel domain and may fail to develop useful skills. Therefore, in both domains, effective instructional design should include appropriate coaching and scaffolding to help all students learn.

In order to provide appropriate scaffolding in introductory physics courses, effective instructional approaches have been based, e.g., on Piaget's model of "optimal mismatch [71-75], Vygotsky's notion of the "zone of proximal development" [76-78], and the preparation for future learning model focusing on "innovation vs. efficiency" by Bransford and Schwartz [79]. Piaget's optimal mismatch model is similar to the "Conceptual Change" model of Posner et al. [31] and suggests that students will benefit if instruction provides a cognitive conflict which makes them understand that there is a mismatch between their naïve, everyday model and what the laws of physics predict in a particular context. Then, students are provided appropriate guidance and feedback for the "assimilation and accommodation" of new ideas consistent with classical physical laws. In line with Piaget's ideas, Posner et al. encourage instructional designers to develop learning activities which allow students to accommodate ideas within the new paradigms with their prior knowledge. They suggest that learning activities should involve creating a state of disequilibrium in students' minds as well as helping them discern anomalies in their knowledge structure, diagnose errors in their thinking, make sense of scientific content by presenting it in multiple representations (verbal, mathematical, graphical, etc.), and translate between representations [31]. The zone of proximal development attributed to Vygotsky is a dynamic zone defined by what a student can accomplish on his/her own at a given time vs. with the help of a guide who is familiar with the student's initial knowledge and targets instruction somewhat above it continuously for effective learning [76-78]. Similarly, Bransford and Schwartz recommend that balanced instruction should include opportunities to learn how to rapidly retrieve and accurately apply appropriate knowledge and skills to solve a problem (efficiency) and to adapt knowledge to new situations (innovation). Students learn most optimally when they follow the "optimal adaptability corridor" in which there are elements of both efficiency and innovation concurrently which helps them be cognitively engaged and prevents them from becoming bored or frustrated [79]. All of these models are synergistic in that one can provide an optimal mismatch by ensuring that instruction is in the zone of proximal development and by designing instructional tasks that are in the "optimal adaptability corridor." Our framework suggests that, similar to instructional strategies in introductory physics, instructional tasks in quantum mechanics which include these types of scaffolding supports and provide sense-making and learning opportunities may help students organize their knowledge coherently and hierarchically while helping them acquire useful skills.

### B. Design of scaffolding supports to help students develop a functional knowledge of quantum mechanics

#### 1. Creation of "a time for telling" to activate prior knowledge and prime students to learn

In order to make the lecture of the instructional design effective, Schwartz and Bransford suggest that instructors create a "time for telling" by first giving students the opportunity to struggle while solving problems and activate relevant prior knowledge before attending the lecture [80]. They suggest that struggling and activating prior knowledge "primes" students to utilize lecture time as a learning opportunity [80]. Due to the facts that many students are inadequately prepared, have unclear goals, and insufficient motivation to excel and there is a paradigm shift from classical to quantum mechanics, our framework suggests that instruction in quantum mechanics should also focus on priming students in order to help all of them learn. Instructional designers can create "a time for telling" [80] by developing research-based learning activities that provide opportunities to activate relevant prior knowledge and make students struggle before lectures in quantum mechanics.

## 2. Research-based active learning tools to improve students' conceptual understanding of quantum mechanics

Research-based active-learning tools such as tutorials [26], peer-instruction [60], group problem-solving [81], and exploiting computers for pedagogical purposes, e.g., the "Just-In-Time Teaching method" [82] are scaffolding tools that help students develop a functional knowledge. They build on students' prior knowledge and explicitly address common difficulties students have in reconciling their naïve mental models with those consistent with the laws of physics. These learning tools give students an opportunity to assimilate and accommodate new ideas while building and organizing their knowledge structure. The guided approach also promotes collaboration and helps students take advantage of each other's strengths and learning styles. Another activity which may help students develop a functional knowledge is asking students in small groups to categorize problems based upon how those problems are solved. While research has shown that students have difficulty categorizing quantum mechanics problems [18], asking them to categorize problems and discuss why certain groupings are better than others may help students look beyond the surface features of the problem, consider the applicability of a physics principle in diverse situations, and "chunk" [42, 83] conceptual knowledge pieces in a hierarchical manner. Based upon the framework, it is likely that research-based active learning strategies like those that have been successful for improving learning in introductory physics (e.g., see Refs. [9, 26, 60, 81, 82, 84]) may be effective for helping students learn quantum mechanics. Indeed, existing research corroborates the implications of the framework to learning quantum mechanics, and research-based pedagogies such as tutorials and peer-instruction tools are proving to be effective in helping students learn quantum mechanics (e.g., see Refs. [11, 20, 85-93]).

Since the patterns of student difficulties in developing expertise in quantum mechanics are analogous to those in introductory mechanics, scaffolding involving research-based active-learning tools that have proven to be successful in introductory courses are likely to be effective in teaching quantum mechanics [10, 11, 85, 86]. Moreover, research has shown that, similar to introductory students, students in quantum mechanics courses can also "co-construct" knowledge when they solve problems with peers [94]. Co-construction of knowledge while working in pairs occurs when neither student in a discussion group can solve the problem individually, but they are able to solve the problem together [95]. In a study by Singh [95], the Conceptual Survey of Electricity and Magnetism [96] was administered to introductory students, both individually and in groups of two. It was found that co-construction occurred in 29% of the cases in which both students had selected an incorrect answer individually. Similarly, a conceptual, multiple-choice test on the formalism and postulates of quantum mechanics was administered to 39 upper-level students, and it was found that co-construction occurred in 25% of the cases in which both students had selected an incorrect answer individually [94]. Thus, even upper-level students benefit from working with peers. Research has already shown that research-based active learning instructional strategies which have been developed for introductory physics such as peer instruction [60] are also effective in the teaching and learning of quantum mechanics [87].

## 3. Explicit guidance to engage students in self-regulatory activities

Similar to introductory physics students, students in upper-level quantum mechanics display varying levels of proficiency in their problem-solving, reasoning, and metacognitive skills. What types of activities may help students improve these skills in upper-level quantum mechanics? Similar to introductory students, upper-level students need explicit guidance to engage in self-regulatory activities. Schoenfeld proposes a few scaffolding methods to explicitly help students develop their problem-solving, reasoning, and metacognitive skills. He suggests that presenting "polished" solutions to the class may hide the problem-solving processes that the instructor engaged in while solving the problem [68]. Instead, Schoenfeld recommends presenting "problem resolutions" in which the instructor models the problem-solving process explicitly by looking through a few examples, making tentative explorations, and asking questions such as "Am I making reasonable progress? Does this seem like the right thing to do?" After the instructor solves the problem, he should assist students in reviewing and evaluating the entire solution, helping them to learn why reflection is an important component of learning from problem solving. This method of teaching may focus students' attention on metacognitive behaviors, even if it is only used in some of the classes. Another method Schoenfeld uses in his mathematics courses is to conduct whole-class discussions of problems while he acts as a scribe or moderator. The entire class decides which methods to pursue while solving a problem, and Schoenfeld asks questions such as "Do things seem to be going pretty well? If not, we might want to reconsider. Are there ideas we want to return to?" These types of sessions are primarily aimed to focus students' attention on their control of learning and self-regulation [68]. Another approach involving reciprocal teaching was used by Reif and Scott [97] to help students learn introductory mechanics. In self-paced computer tutorials called Personal Assistants to Learning (PALs), computers and students took turns to help each other solve physics problems. This approach was found to be effective in improving students' self-regulatory skills.

Another instructional approach that explicitly encourages students to view problem-solving as a learning opportunity is requiring students to correct their mistakes on homework, quizzes, and exams [43, 44, 98]. Introductory students who self-diagnosed their mistakes performed significantly better on a follow-up exam [44]. Based upon our framework, students taking quantum mechanics may not automatically learn from their mistakes and may also benefit from self-diagnosing their mistakes on homework, quizzes, and exams similar to introductory students. Students taking quantum mechanics should be rewarded appropriately for self-diagnosis activity; otherwise, they may not engage with the material deeply. In fact, research has already shown that when students in quantum mechanics were given grade incentives to fix their mistakes on a midterm exam, they did significantly better on similar final exam problems than students who were not given a grade incentive to fix their mistakes on the midterm exam [99].

### 4. Instructional strategies to improve students' epistemological views

Developing a functional knowledge is closely connected to appropriate epistemological views of the subject matter. What types of instructional strategies can help improve students' epistemological views? Based upon the framework, similar to instructional strategies which improve students' epistemological views in introductory mechanics, students' epistemological views about learning quantum mechanics can be improved if instructional design focuses on sense-making and learning rather than on memorization of facts and accepting the instructor as the authority. These effective instructional strategies should include students working with peers to make sense of the material and providing problems in contexts that are interesting and appealing to students. Both formative assessments (e.g., homework, in-class conceptual questions, group problem-solving, etc.) and summative assessments (e.g., exams) should include context-rich problems and sense-making problems to evaluate whether students can apply quantum mechanical principles to a real-world setting. Otherwise, students will continue to "game" exams by successfully solving complex algorithmic problems involving, e.g., solutions of the time-independent Schrödinger equation with complicated boundary conditions and potential energies without having developed a functional understanding of quantum mechanics. Additionally, similar to instructors of introductory physics, instructors of quantum mechanics should choose their terminology carefully and be consistent to avoid negatively impacting student learning. For example, it is important that students become aware of the difference between "doing" quantum mechanics and "understanding" quantum mechanics (as alluded to by Feynman). In particular, the curriculum should help students understand that while there are many interpretations of quantum mechanics, there are interpretations with well-established postulates and procedures for predicting quantum mechanical outcomes in diverse situations (e.g., the Copenhagen interpretation). Instructors should guide students to make sense of these postulates and procedures to evaluate outcomes of experiments. These activities may further improve students' epistemological views about quantum mechanics, encourage them to engage in self-regulatory activities, and help them organize their knowledge structure of quantum mechanics.

### 5. Types of assessment to encourage students to develop a functional understanding

Mathematically skilled students in a traditional introductory physics course focusing on mastery of algorithms without conceptual understanding can "hide" their lack of conceptual knowledge behind their mathematical skills [60]. However, their good performance on algorithmic physics problems does not imply that they have engaged in self-regulation throughout the course or have built a hierarchical knowledge structure. In fact, most physics faculty, who teach both introductory and advanced courses, agree that the gap between conceptual and quantitative learning gets wider in a traditional physics curriculum from the introductory to advanced level [100]. Therefore, students in a traditionally taught and assessed quantum mechanics course can "hide" their lack of conceptual knowledge behind their mathematical skills even better than students in introductory physics. Based upon the framework, research on student learning in introductory physics suggests that closing the gap between conceptual and quantitative problem-solving by assessing both types of learning is essential in helping students in quantum mechanics develop functional knowledge [10]. Interviews with faculty members teaching upper-level quantum mechanics suggest that some assign only quantitative problems in homework and exams (e.g., by asking students to solve the time-independent Schrödinger equation with complicated boundary conditions) because they think students will learn the concepts on their own [100]. Nevertheless, a majority of students may not learn much about quantum mechanics concepts including the formalism unless course assessments value conceptual learning, sense-making, and the building of a robust knowledge structure. Therefore, to help students develop a functional knowledge of quantum mechanics, formative and summative assessments should emphasize the connection between conceptual understanding and mathematical formalism. Since assessment drives learning (i.e., students will learn what they are tested on), formative assessment can be an effective way to coach and scaffold students [101]. Students who are assessed on both conceptual and quantitative understanding of quantum mechanics throughout the semester are more likely to acquire a functional knowledge similar to the findings from research in introductory physics teaching and learning.

Instructors should also assess students' self-regulation. One way this can be done is by requiring students to explicate their reasoning while solving problems. Research has shown that students in introductory physics (and other introductory science courses) who articulate their reasoning, or "self-explain," while studying worked physics examples can detect conflicts in their knowledge structure and they can be coached explicitly on effective self-explanation processes while learning on their own [43, 98, 102]. It is recommended that instructors provide students with prompts that encourage students to detect conflicts [43, 98]. Based upon our framework, students in quantum mechanics courses display difficulties in self-regulation similar to introductory physics students, so quantum mechanics instructors should also assess students' explication and reasoning while solving problems. Further, assessments should evaluate students' self-regulatory skills by considering the consistency and sense-making in their responses. These types of assessments may explicitly focus on coaching and scaffolding student learning to help students self-regulate and engage in sense-making while solving quantum mechanics problems.

### C. Concluding remarks

Consistent with the framework, the existing research-based instructional tools for helping students learn quantum mechanics that are inspired by similar tools for introductory physics are already proving to be effective [10, 11, 65, 85-87]. As further research is conducted in quantum mechanics teaching and learning, we will learn more about the patterns of difficulties and the nature of expertise so that the framework presented here can be refined further.

## ACKNOWLEDGEMENTS

We thank the National Science Foundation for awards PHY-0968891 and PHY-1202909. We also thank F. Reif, R. P. Devaty and all members of the physics education research group at the University of Pittsburgh for their valuable feedback.